\documentclass[fleqn,usenatbib]{mnras}

\usepackage[T1]{fontenc}

\DeclareRobustCommand{\VAN}[3]{#2}
\let\VANthebibliography\thebibliography
\def\thebibliography{\DeclareRobustCommand{\VAN}[3]{##3}\VANthebibliography}

\usepackage{graphicx}	% Including figure files
\usepackage{amsmath}	% Advanced maths commands
\usepackage{amssymb}	% Extra maths symbols

\newcommand\laemabs{M}
\newcommand\vol{V}

\newcommand\pakidge{\texttt{galcv}}
\newcommand\Nobs{N_{\rm obs}}
\newcommand\Nexp{N_{\rm exp}}

\newcommand\Nexpavg{N_{\rm avg}}
\newcommand\Fsurvavg{\bar{F}_{\rm surv}}
\newcommand\Fsurv{F_{\rm surv}}
\newcommand\Nvis{N_{\rm avg,vis}}
\title[LAEs in Ionized Bubbles]{Lyman-$\alpha$ Emitters in Ionized Bubbles: Constraining the Environment and Ionized Fraction}
\author[Trapp, Furlanetto, \& Davies]{
A.C. Trapp,$^{1}$\thanks{E-mail: atrapp@ucla.edu}
Steven R. Furlanetto,$^{1}$ \& Frederick B. Davies$^{2}$
\\
$^{1}$Department of Physics and Astronomy, University of California Los Angeles, CA, 90095-1562, USA \\
$^{2}$Max-Planck-Institut f\"{u}r Astronomie, K\"{o}nigstuhl 17, 69117 Heidelberg, Germany}

\date{Accepted XXX. Received YYY; in original form ZZZ}

\pubyear{2022}

\begin{document}
\label{firstpage}
\pagerange{\pageref{firstpage}--\pageref{lastpage}}
\maketitle

\begin{abstract}
Lyman-alpha emitters (LAEs) are excellent probes of the reionization process, as they must be surrounded by large ionized bubbles in order to be visible during the reionization era. Large ionized regions are thought to correspond to over-dense regions and may be
%(or contain) 
protoclusters, making them interesting test-beds for early massive structures.
%For these reasons, 
Close associations containing several LAEs are often assumed to mark over-dense, ionized bubbles.
Here, we develop the first framework to quantify the ionization and density fields of high-$z$ galaxy associations.
%, and we study how these galaxies can be used to learn about the progress of reionization. 
We explore the interplay between \textit{(i)} the large-scale density of a survey field, \textit{(ii)} Poisson noise due to the small number density of bright sources at high redshifts ($z \sim$ 7), and \textit{(iii)} the effects of the ionized fraction on the observation of LAEs.
We use Bayesian statistics, a simple model of reionization, and a Monte-Carlo simulation to construct a more comprehensive method for calculating the large-scale density of LAE regions than previous works.
We find that Poisson noise has a strong effect on the inferred density of a region 
%if there are a small number of sources used to calculate the density 
and show how the ionized fraction can be inferred. We then apply our framework to the strongest association yet identified: Hu et al. (2021) found 14 LAEs in a volume of $\sim$50,000 cMpc$^3$ inside the COSMOS field at $z \sim7$. We show that this is most likely a 2.5$\sigma$ over-density 
%(despite there being nine times more visible LAEs than expected in that region) 
inside of an ionized or nearly ionized bubble. We also show that this LAE association implies that the global ionized fraction is $\bar{Q} = 0.60^{+0.08}_{-0.09}$, within the context of a simple reionization model. 
%Our approach provides the first step in quantifying the implications of specific LAE associations for reionization.  
\end{abstract}


%%%%%%%%%%%%%%%%%%%%%%%%%%%%%%%%%%%%%%%%%%%%%%%%%%

%%%%%%%%%%%%%%%%% BODY OF PAPER %%%%%%%%%%%%%%%%%%

%%%%%%%%%%%%%%%%%%%%%%%%%%%%%%%%%%%%%%%%%%%%%%%%%%%%%%%%%%%%%%%%
%%%%%%%%%%%%%%%%%%%%%%%%%%%%%%%%%%%%%%%%%%%%%%%%%%%%%%%%%%%%%%%%
\section{Introduction}

%\textbf{Background:}
The Cosmic Dawn is about to be explored as never before, and a spotlight shines on reionization. The search for rare, large ionized regions and the first large-scale objects like protoclusters is of particular interest \citep{Ouchi2005,Calvi2019,Tilvi2020,Jung2020,Hu2021,Endsley2021}. These extreme regions may serve as test-beds for the physics of early galaxy and cluster formation.
By measuring the large-scale densities of these regions, we can learn about the assembly history of galaxy clusters and other large, rare objects \citep[e.g.][]{Chiang2017}.
Also, feedback on the scale of these extreme regions, driven by underlying large-scale structure, is likely to affect the galaxy population strongly during reionization \citep{Thoul1996, Iliev2007, Noh2014}.

Lyman-alpha emitters (LAEs) are exciting probes of reionization (as well as galaxy formation). The young galaxies present in the reionization era are likely to have large intrinsic Lyman-$\alpha$ luminosities \citep{Partridge1967,Ouchi2020}, but those photons are subject to substantial absorption by the intergalactic medium (IGM) before reionization is complete, thanks to the enormous optical depth of remaining neutral islands \citep{Gunn1965, MiraldaEscude1998}. We therefore expect to see a decline in the abundance of LAEs as we penetrate further into the reionization era, making these galaxies an effective probe of the ionization state \citep{Madau2000, Haiman2002, Santos2004}.  

The number counts of LAEs evolve rapidly above $z \sim 5$, which may in part be attributable to reionization \citep{Malhotra2004, Kashikawa2011, Itoh2018}. However, the evolution of the
number density of LAEs is difficult to disentangle from the overall evolution of the galaxy abundances. For that reason, attention has shifted (when possible) to  more sophisticated ways to leverage LAEs. One avenue is to measure the fraction of photometrically-selected galaxies with Lyman-$\alpha$ lines \citep{Pentericci2011, Stark2011, Ono2012}. These studies have typically found that the Universe must have a substantial neutral fraction at $z \sim 7$, though the inferences depend on the reionization model \citep{Mason2018, Morales2021}. 

Another exciting prospect is that the inhomogeneous ionization field will modify the spatial distribution of LAEs. In order for their Lyman-$\alpha$ lines to survive, the host galaxy must be embedded in a large ionized region, which, in most reionization models, corresponds to an over-dense region
\citep{Furlanetto2004, Wyithe2005}. Galaxies outside of such large regions may produce Lyman-$\alpha$ photons, but they will not survive the neutral IGM. Thus the ionization field modulates the clustering of LAEs
\citep{Furlanetto2006, McQuinn2007}. Unfortunately, these clustering measurements are challenging and require a large number of sources (e.g., \citealt{Yoshioka2022}). 

A particularly interesting aspect of this modulation is that it exaggerates the existing clustering: galaxies in over-dense regions (which host large ionized bubbles) will remain visible, while even galaxies inside modest over-densities will become invisible during the early stages of reionization. This suggests that focusing on identifying rare ionized regions with LAEs can be a powerful probe of reionization (e.g., \citealt{Mesinger2008}). 

Meanwhile, surveys for LAEs, typically with narrowband filters, are now common, with large enough areas that unusual regions can be found. Of particular interest for us, several teams have discovered candidate ionized bubbles hosting apparently significant over-densities of LAEs 
(e.g., \citealt{Tilvi2020,Jung2020,Hu2021,Endsley2021}).

Perhaps the most compelling such region has recently been identified by \citet{Hu2021}, who
found 14 LAEs in a region with volume $\vol \sim50,000$ cMpc$^3$ at $z =$ 6.93 ($\sim200$ pMpc$^3$) 
within a survey that encompassed approximately 118 total independent volumes of that size \citep[see also][]{Wold2022}.
Using the total number of LAEs in these larger fields, the expected number to be found in this over-dense volume is just $\Nexpavg = 1.5\pm 0.1$. This is thus a clear over-density in the LAE counts (nine times more sources than expected), and hence very likely a large ionized bubble.
%\textbf{The Problem:}
However, to date such inferences are purely qualitative, without attempting to transform observed features in the galaxy distribution into a quantitative constraint on the underlying ionization and density fields. 

Finding and analysing these regions, and then connecting the visible sources to the total population of galaxies that are ionizing the Universe, will be challenging on many fronts.
The bulk of ionizing photons, at least at early times, are very likely produced by galaxies that we cannot see with HST \textit{nor} JWST \citep[see e.g.,][]{Behroozi2015,Furlanetto2017}. 
In a partially-ionized Universe, even strong Ly$\alpha$ lines can be obscured by intervening neutral hydrogen, leaving only a small number of the brightest LAEs in the most ionized regions visible. The observed number density of these LAEs is then strongly affected by Poisson noise.
Poisson noise is uncorrelated between magnitude bins, meaning an over-density of a few bright LAEs does not necessarily mean an over-density in the much-larger underlying population of faint or obscured sources. 
In other words, determining the large-scale density of a region from a small number of LAEs requires a simultaneous treatment of \textit{(i)} cosmic variance and the corresponding galaxy bias of those LAEs, \textit{(ii)} Poisson noise, and \textit{(iii)} the effects of a partially ionized Universe on the visibility of LAEs.

The first two of these points are more general than LAEs: programs to find ``protoclusters" or other unusual environments from galaxy distributions suffer from Poisson noise as well. Such efforts to identify protoclusters are useful for tracing the history of the most massive structures in the Universe today
\citep{Ouchi2005, Trenti2012, Chiang2017,  Calvi2019,Hu2021}, but the associations have also been largely qualitative. In particular, the probability that a galaxy overdensity will virialize by the present day, and on what scale that virialization will occur, has not generally been quantified. 

%\textbf{Our Purpose:}
In this paper, we develop the first quantitative, statistically robust framework to infer the underlying density and ionization environment of observed galaxy associations during the reionization era. This framework can be applied both to LAEs (which probe both the ionization state and the density) and to other surveys (which probe only the density). 

In section~\ref{lae_sec:poissondensity}, we construct an analytic form of the posterior for the large-scale density of a region, 
ignoring any effects of inhomogeneous reionization. 
We apply this method to the \citet{Hu2021} region 
and another, less extreme over-density \citep{Endsley2022} 
in section~\ref{lae_sec:resultsIonized}.
We then expand our method to the partially-ionized case using a simple model of reionization and a Monte-Carlo (MC) simulation in section~\ref{lae_sec:methodspartial}. This MC simulation also has the ability to constrain the ionized fraction of the Universe (in the context of our simple model of reionization).
In section~\ref{lae_sec:resultsPartiallyIonized}, we apply the MC simulation method to the same
regions from \citet{Hu2021} and \citet{Endsley2022}, obtaining a new measurement of its density 
and local ionization field as well as a constraint on the ionization fraction of the Universe.

%\textbf{Cosmology:}
We use the following cosmological parameters: $\Omega_m = 0.308$, $\Omega_\Lambda=0.692$, $\Omega_b=0.0484$, $h=0.678$, $\sigma_8=0.815$, and $n_s=0.968$, consistent with recent Planck Collaboration XIII results \citep{PlanckCollaboration2016}.
%%%%%%%%%%%%%%%%%%%%%%%%%%%%%%%%%%%%%%%%%%%%%%%%%%%%%%%%%%%%%%%%
%%%%%%%%%%%%%%%%%%%%%%%%%%%%%%%%%%%%%%%%%%%%%%%%%%%%%%%%%%%%%%%%

\section{Inference of the local density field from galaxy associations}
%\section{Folding Poisson noise into density measurement}
\label{lae_sec:poissondensity}

We begin by ignoring reionization and just imagining inferring the underlying dark matter density of a field with some set of observed sources. 
Take the average number density of observable galaxies in a survey at a given redshift to be $n_{\textrm{avg}}$. In a volume $\vol$, one would expect to find $\Nexpavg = \vol \cdot n_{\textrm{avg}}$ galaxies. Now, assume one completes an observational campaign of size $N_{\textrm{vols}} \times \vol$, finding that the most dense region of size $\vol$ has
$N_{\textrm{obs}} \gg \Nexpavg$. What can we infer about that region? Does the ratio of $N_{\textrm{obs}}/\Nexpavg$ carry through to those sources below the magnitude limit of the survey?
Can we measure the excess amount of dark matter in this region using a bias function and infer whether the region will collapse into a cluster by $z =$ 0? To answer these questions, one must consider Poisson noise and cosmic variance jointly.

In the gaussian approximation of cosmic variance, the expected number of sources in some region $\vol$ with \textit{linearized} relative density $\delta = (\rho-\bar{\rho})/\bar{\rho}$, is
\begin{equation}\label{lae_eq:conditional1}
    \Nexp = \Nexpavg (1 + \delta \cdot b(\vol)),
\end{equation}
where $b(\vol)$ is the \textit{bias} of those sources\footnote{The bias also typically depends on the luminosity or mass of the objects being considered. However, for a set population defined by
an intrinsic luminosity function and
magnitude range, there is an effective bias value for that population.} \citep{Mo1996}. However, when one observes a region of density $\delta$ and corresponding $\Nexp$, one does not always find $\Nobs = \Nexp$. We assume that the observed number is drawn from the Poisson distribution with $\lambda = \Nexp$; explicitly,
\begin{equation}\label{lae_eq:poisson}
    P(\Nobs|\lambda=\Nexp) = \frac{\Nexp^{\Nobs}e^{-\Nexp}}{\Nobs!}.
\end{equation}
Simulations show that, at least for some galaxy populations, the variance becomes super-Poisson in the nonlinear regime (e.g., \citealt{Ahn2015,Friedrich2018, Gruen2018, Friedrich2022}), but we do not attempt to model such effects. We note that any increase in the variance would only weaken evidence for strongly over-dense regions, so our choice is conservative in this sense.

One can then infer the value of $\Nexp$ given $\Nobs$ using Bayes' theorem,
\begin{equation}
    p(\Nexp|\Nobs) \propto P(\Nobs|\lambda = \Nexp)p(\lambda = \Nexp),
\end{equation}
where $p(\lambda = \Nexp)$ is the prior on $\Nexp$. From equation~(\ref{lae_eq:conditional1})
and the fact that the cosmological density field is a Gaussian (at least in the linear approximation appropriate on large scales), the prior $p(\lambda = \Nexp)$ is a Gaussian centered at $\Nexpavg$ with standard deviation $\sigma_{\Nexp} = \Nexpavg \cdot b \cdot \sigma_\delta$,
\begin{equation}\label{lae_eq:gaussian}
    p(\Nexp) = \frac{1}{\sigma_{\Nexp} \sqrt{2\pi}} \exp \left[ -\frac{1}{2} \left( \frac{\Nexp - \Nexpavg}{\sigma_{\Nexp}}\right)^2 \right],
\end{equation}
where $\sigma_\delta$ is the r.m.s. fluctuation of $\delta$ 
%on a scale of 66 x 30 x 26 cMpc$^3$, the size of the \citet{Hu2021} volume 
for the geometry matching the survey volume \citep[see][for calculations of non-spherical density fluctuations]{Newman2002, Stark2007, Munoz2010, Robertson2010, Trapp2020}.
Multiplying equations~(\ref{lae_eq:poisson}) and~(\ref{lae_eq:gaussian}), taking the natural log, and dropping terms that do not depend on $\Nexp$ gives the log likelihood. Changing the inferred quantity to the underlying dark matter density $\delta$ via 
%\begin{equation}
%    {\textrm{ln}} p(\Nexp|\Nobs) \propto \Nobs{\textrm{ln}}\Nexp - \Nexp - \frac{1}{2}\left( \frac{\Nexp - \Nexpavg}{\sigma_{\Nexp}}\right)^2.
%\end{equation}
%Substituting in 
equation~(\ref{lae_eq:conditional1}), we find
%to change variables to $\delta$,
\begin{equation}\label{lae_eq:deltapost}
    {\textrm{ln}} \ p(\delta|\Nobs) \propto \Nobs {\textrm{ln}}(1 + \delta \cdot b) - \Nexpavg \cdot \delta \cdot b - \frac{1}{2}\left( \frac{ \delta}{ \sigma_\delta}\right)^2.
\end{equation}

The above equation provides the full posterior of the region's dark matter density given an observed number of sources $\Nobs$, an average expected number of sources $\Nexpavg$, a bias value $b$ for those sources, and knowledge of the r.m.s. fluctuation in the linear dark matter density field $\sigma_\delta$. 

If there is uncertainty in one of those parameters -- say the bias value $b$ -- its probability distribution $p(b)$ can be marginalized over in the following way:
\begin{equation}
    p(\delta|\Nobs) =\int p(\delta, b|\Nobs) db = \int p(\delta|b,\Nobs)p(b)db. 
\end{equation}

So far, we have imagined a simple experiment in which only the total number of sources is known. We will restrict ourselves to this simple case in this paper, but it is easy to extend the formalism to more sophisticated experiments. In a more ideal case, the average luminosity function of these sources $\Phi_{\textrm{avg}}$ is known, with corresponding local luminosity function $\Phi_{\textrm{loc}}(\laemabs) = \Phi_{\textrm{avg}}(\laemabs) (1 + \delta \cdot b(\laemabs,\vol))$, where $\laemabs$ is the absolute magnitude of the sources. In this case, with $\Nobs$ observed sources, each with $\laemabs_j$, the posterior becomes
\begin{eqnarray}\label{lae_eq:deltapostLF}
    {\textrm{ln}} \  p(\delta|\Nobs) & \propto & - \Nexpavg \cdot \delta \cdot b - \frac{1}{2}\left( \frac{ \delta}{ \sigma_\delta}\right)^2 + \\
    & & \sum_j^{\Nobs} \Big[ \textrm{ln}\Phi_{\textrm{avg}}(M_j) +  \textrm{ln}\left[ 1+\delta  b(\laemabs_j,\vol)\right] \Big].
\end{eqnarray}

\begin{figure*}
    \centering
    \includegraphics[width=0.485\textwidth]{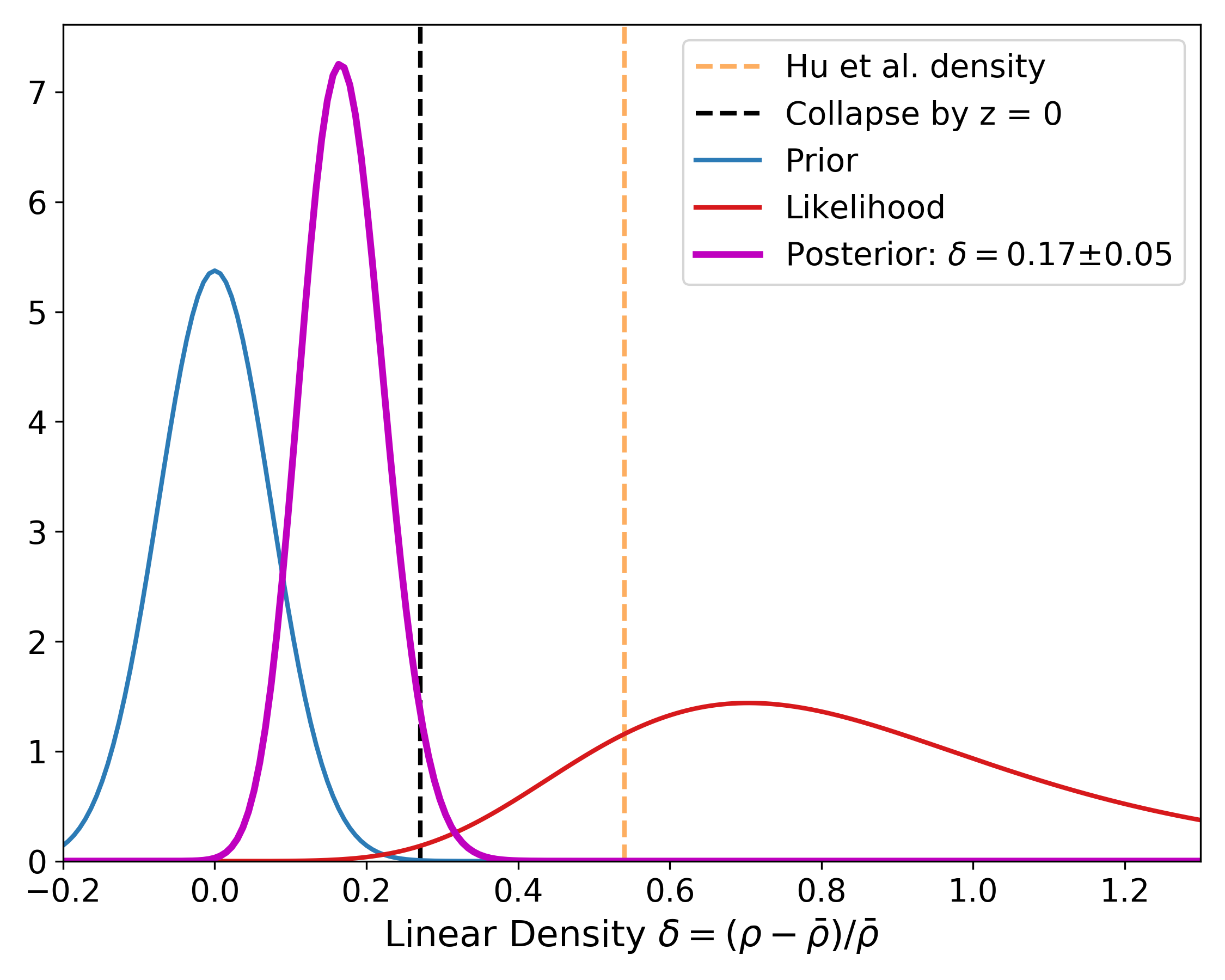}
    \includegraphics[width=0.485\textwidth]{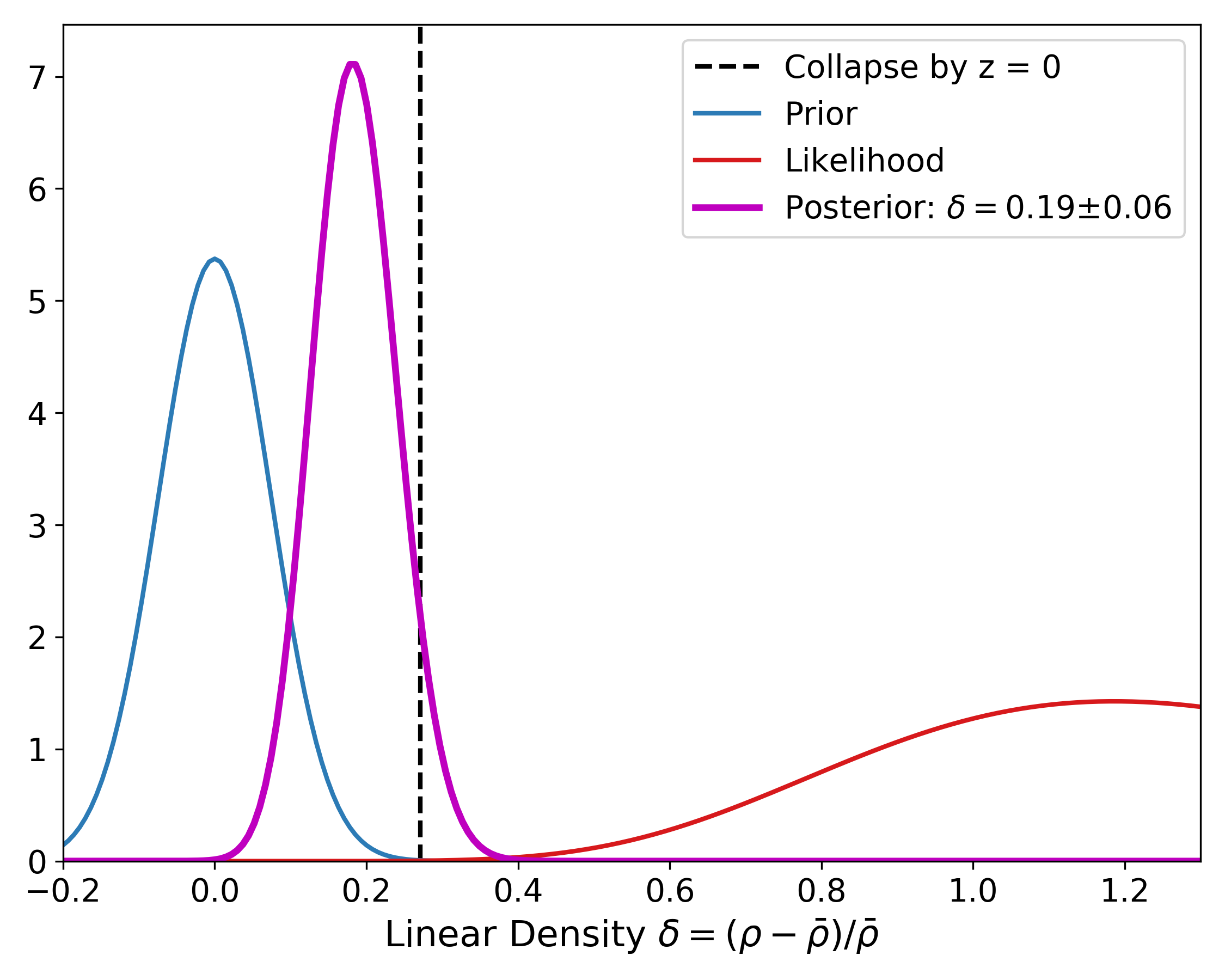}
    \caption{
    The inferred density  for the \citet{Hu2021} volume, assuming no effect from reionization.  \emph{Left:} The \textit{magenta curve} shows the posterior of the linear density, which has 68.27\% credible interval
    $\delta = 0.17 \pm 0.05$, taking the average source density from the COSMOS field alone.
    The vertical dashed \textit{black} line indicates the approximate density required for this region to virialize by $z = $ 0, $\delta_{\textrm{pc}} = 0.27$, which is disfavored by the posterior. The vertical dashed \textit{orange} line shows the density estimate of \citet{Hu2021}.
    The \textit{red and blue} curves show the likelihood and prior, respectively;
    we are in the prior-dominated regime, given the likelihood's large breadth.
    The likelihood and prior are very far apart, indicating that this volume is a rare find; we show in the text that
    the probability of finding at least one such region in a larger survey of 20 similar volumes (the COSMOS field) is 0.11\%. \emph{Right:} Same, but using the $\Nexpavg$ value from all 4 LAGER fields. The inferred density increases from $\delta = 0.17 \rightarrow 0.19$. The probability of finding at least one such region in a larger survey of 118 similar volumes (all 4 LAGER fields) drops to 0.01\%.}
    \label{lae_fig:HuPost_COSMOS}
\end{figure*}

%\begin{figure}
%    \centering
%    \includegraphics[width=0.485\textwidth]{laefig_ionized_denpost_linear_LAGERNavg.png}
%    \caption{The same as Figure~\ref{lae_fig:HuPost_COSMOS}, but using the $\Nexpavg$ value from all 4 LAGER fields. The inferred density is increased from $\delta = 0.17 \rightarrow 0.19$. The probability of finding at least one such region in a larger survey of 20 similar volumes (the COSMOS field) drops to 0.002\%.}
%    \label{lae_fig:HuPost_LAGER}
%\end{figure}

%%%%%%%%%%%%%%%%%%%%%%%%%%%%%%%%%%%%%%%%%%%%%%%%%%%%%%%%%%%%%%%%
%%%%%%%%%%%%%%%%%%%%%%%%%%%%%%%%%%%%%%%%%%%%%%%%%%%%%%%%%%%%%%%%
\section{Applying the density inference framework}
%\section{Results for fully-ionized Universe}
\label{lae_sec:resultsIonized}

\subsection{An apparent extreme overdensity}

In \citet{Hu2021}, 14 LAEs were found in a region with volume $\vol = 66 \times 30 \times 26$ cMpc$^3$ at $z =$ 6.93 ($\sim200$ pMpc$^3$), embedded within a larger survey field with approximately 20 volumes of the same size. 
There are also three other similar fields \citep{Hu2021,Wold2022}, the total volume of which is 118 times the volume of the over-dense region.
Henceforth, for convenience we will refer to this specific over-dense volume of 14 sources as the ``\citet{Hu2021}  volume'', and in general we will call a group of galaxies found at high-$z$ an ``association''. Such associations are often referred to as protoclusters, but we reserve that term to refer to systems that will virialize by the present day.

For simplicity, to constrain the properties of this region we will use priors on the observed parameters ($\Nexpavg$ and the bias $b$) determined by \citet{Hu2021}. In principle, these quantities could be determined by an underlying theoretical model. The expected number of LAEs to be found in this volume -- using the number density in the larger COSMOS field -- is
$\Nexpavg = 2.2\pm 0.3$ (not accounting for redshift-space distortions\footnote{In \citet{Hu2021}, the authors include a redshift-space distortion correction to the density of their region. We do not include such a correction here as our measured density values are much lower, corresponding to a redshift-space distortion effect of less than 10\%, well below our other uncertainties.}).
\citet{Hu2021} calculated a galaxy over-density of $\delta_g = 5.11^{+2.06}_{-1.70}$ for the region with 14 sources. Combining this with a bias value of $b_{\textrm{real}} = 4.54 \pm 0.63$ and a redshift-space distortion correction $C=0.79$ via the equation $1 + b_{\textrm{real}} \delta_{\textrm{real}} = C(1+\delta_g)$, they found the dark-matter over-density to be $\delta_{\textrm{real}} = 0.87$. Converting this real over-density to a linear over-density via \citet{Mo1996} gives $\delta = 0.54$.
Given the r.m.s. fluctuation in the linear density field for the \citet{Hu2021} volume $\sigma_R=$0.074, this corresponds to a $7.3$ $\sigma_R$ over-density, a very large excursion.

We adopt the bias value used in \citet{Hu2021}, though convert it to a linear bias value $b = 7.31\pm1.02$ via $(1+\delta_{\textrm{real}}\cdot b_{\textrm{real}})=(1+\delta \cdot b)$.
With these parameters, using equation~(\ref{lae_eq:deltapost}) and marginalizing over the uncertainty in $b$ and $\Nexpavg$, we plot the posterior of the density of this region in the left panel of Figure~\ref{lae_fig:HuPost_COSMOS} in \textit{magenta}, finding a density and 68.27\% credible interval of $\delta = 0.17\pm0.05$, much lower than estimated in \citet{Hu2021} (shown by the vertical dashed \textit{orange} line).
Given the r.m.s. density fluctuation in the linear density field for the \citet{Hu2021} volume $\sigma_R=$0.074, this corresponds to a $2.3 \pm 0.7$ $\sigma_R$ over-density.
The Figure also shows the likelihood in \textit{red} (eq.~\ref{lae_eq:deltapost} without the final term, still marginalized over $b$ and $\Nexpavg$) and the prior on the density in \textit{blue}.

In \citet{Wold2022}, COSMOS and 3 additional fields are analysed, giving a total of 174 sources. This changes the expected number of sources to find in the \citet{Hu2021} volume to $\Nexpavg = 1.5 \pm 0.1$. This would imply an even greater over-density, and the right panel of Figure~\ref{lae_fig:HuPost_COSMOS} shows the inferred results. We find a density and 68.27\% credible interval of $\delta = 0.19\pm0.06$, now  a $2.6 \pm 0.8$ $\sigma_R$ over-density, only slightly larger than the previous case.

Our posteriors are prior-dominated, as the likelihoods are much broader than the prior, even with 14 observed sources.
The vertical dashed \textit{black} lines indicate the approximate density required for this region to collapse into a cluster by $z = $ 0, $\delta_{\textrm{pc}} = 0.27$ (assuming the spherical collapse model), a situation that is disfavored by both posteriors. 
If we do not marginalize over the uncertainty in $b$ and $\Nexpavg$, the posterior does not change, with $\delta = 0.19\pm0.06$. This is not surprising, given we are in the prior-dominated regime. If we significantly change the bias value to $b = 5.0 \pm 0.7$ -- the bias for a $\sim4 \times 10^{11} M_\odot$ halo -- the density inference surprisingly shifts to a \textit{lower} value, $\delta = 0.17\pm0.06$, due to the fact that the likelihood, while shifted to a higher average value, becomes considerably more spread out at lower bias values.

In both panels of Figure~\ref{lae_fig:HuPost_COSMOS}, the likelihood and prior are very far apart, indicating that this volume is a rare find. In fact, the probability for a single region of this volume to host 14 \textit{or more} sources given the estimates of $\Nexpavg$, $b$, $\sigma_R$ (including the uncertainties in these quantities), and Poisson noise, is only 0.006\%
when using $\Nexpavg = 2.2 \pm 0.3$, calculated from the COSMOS field alone. When using $\Nexpavg = 1.5 \pm 0.1$ from all 4 LAGER fields, this probability drops to 0.0001\%, about one-in-a-million.
The probability of finding at least one such region in a larger survey of 20 (COSMOS) similar volumes is
0.11\% for $\Nexpavg = 2.2 \pm 0.3$ and 0.01\%\footnote{These are not perfect estimates, as \citet{Hu2021} identified the over-density \emph{after} their survey was completed, so they had freedom to specify its size. The real probability may therefore be somewhat higher, but it is impossible to quantify.} for $\Nexpavg = 1.5 \pm 0.1$ in 118 (all 4 LAGER) larger fields. This is surprising enough that one might ask if another factor might change the probability -- such as a patchwork of ionized bubbles modulating the LAE surface density, as we will explore below. 

The inferred density of this region is a sobering reminder of the difficulty of identifying unusual structures in realistic circumstances: we have analyzed the Hu region precisely because it appears so much more extreme than many other galaxy associations at $z > 6$. While it is a clear over-density, it is most likely \emph{not} going to virialize -- at least across the entire volume -- by the present day. In part, this is because of the scale of the region, which is far larger than a present-day galaxy cluster. It is still possible that a true protocluster that will virialize by the present day is buried within the Hu volume, but the entire volume likely corresponds to a more modest overdensity. Associations of just a few objects, even if they are very luminous, will be even less persuasive (see section~\ref{lae_sec:Endsley}), because they are even more subject to Poisson fluctuations.

We note that protocluster over-densities have been identified with higher confidence at lower redshifts, but these are generally in a regime where many galaxies are identified (so that Poisson fluctuations are small). For example, \citet{Topping2018} identified two such associations in the SSA22 field at $z \sim 3$. Because the properties of these galaxies are well-constrained through clustering measurements, the galaxy bias could be estimated reliably, and the large number ($\sim$100) of galaxies in the associations meant that the number counts could be transformed directly into the dark matter density. Unfortunately, the clustering of high-$z$ LAEs is still difficult to measure, especially for the \citet{Hu2021} sample, and the number counts are so small that Poisson fluctuations cannot be ignored. 

We emphasize that in this section we have ignored any effect incomplete reionization may have on the LAE visibility. But the significant over-density in the Hu volume already suggests that its ionization environment could be very different from the average. 
Using the best-fit luminosity function from 
\citet{Bouwens2021}
and the UV galaxy bias function from \citet{Trapp2020}, integrating down to $M_{\textrm{UV}} = -13$, we estimate that this region has an excess in UV light production of
$70 \pm 23$\% over the average. We will consider the implications of this difference in a later section.

%%%%%%%%%%%%%%%%%%%%%%%%%%%%%%%%%%%%
\subsection{A second over-density}\label{lae_sec:Endsley}

\begin{figure}
    \centering
    \includegraphics[width=0.485\textwidth]{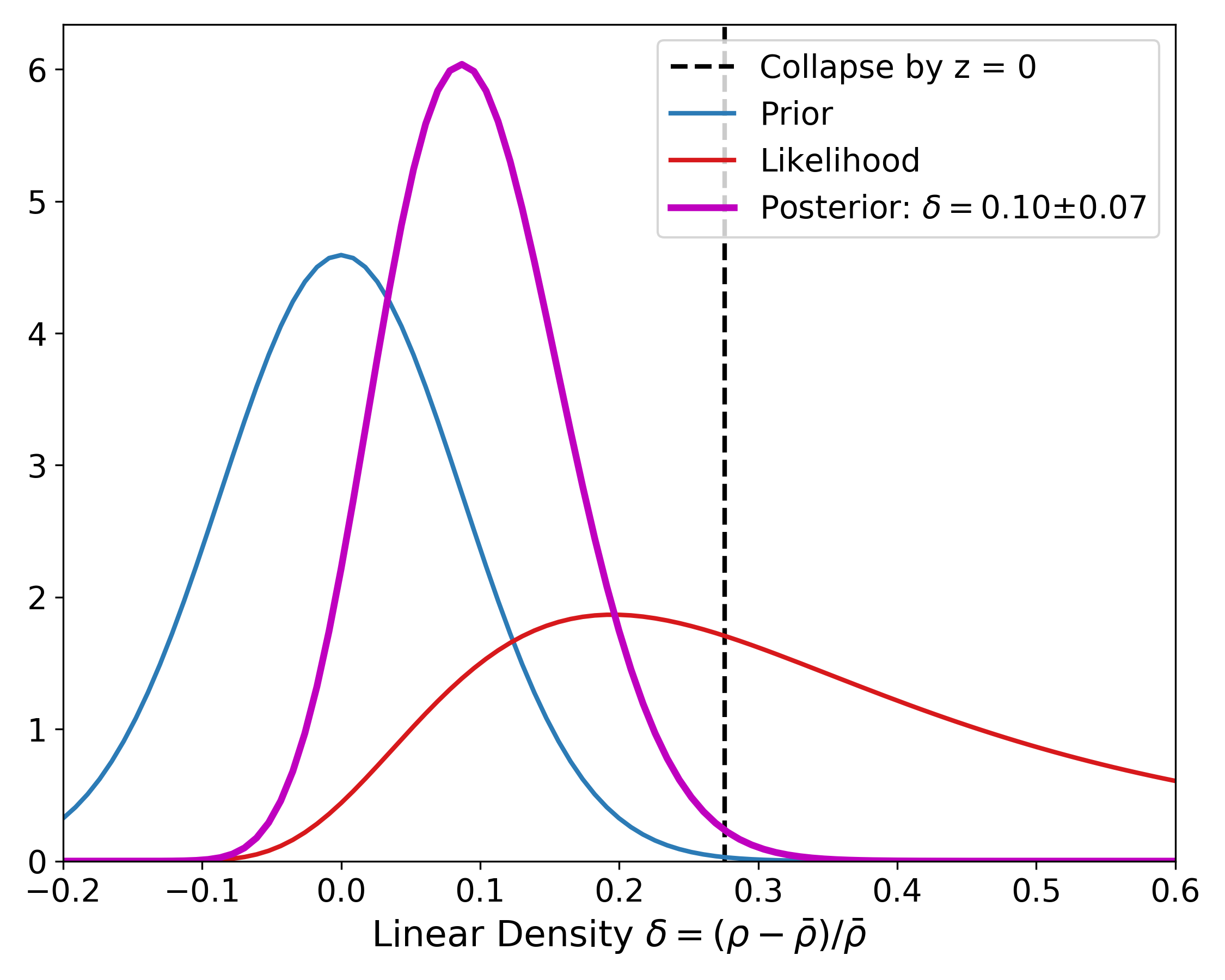}
    \caption{
    The inferred density  for the \citet{Endsley2022} volume, assuming no effect from reionization.  The \textit{magenta curve} shows posterior of the linear density, which has 68.27\% credible interval
    $\delta = 0.10 \pm 0.07$.
    The vertical dashed \textit{black} line indicates the approximate density required for this region to collapse into a cluster by $z = $ 0, $\delta_{\textrm{pc}} = 0.27$, which is disfavored by the posterior.
    The \textit{red and blue} curves show the likelihood and prior, respectively.}
    \label{lae_fig:EndsleyPost}
\end{figure}

\citet{Endsley2022} found an over-density of 6 LAEs in a region of size $3.5 \times 4.8 \times 8.3$ pMpc ($\sim$ 140 pMpc$^3$) at $z \simeq 6.7 - 6.9$ (in an association tentatively first identified in \citealt{Endsley2021}). They calculate the average number of sources expected in such a region to be $\left<N\right> = 2.0^{+0.9}_{-0.7}$ from the \citet{Harikane2022} luminosity function. Using these values, and the bias value of $b = 7.31\pm1.02$, we measure the linear matter density to be $\delta = 0.10\pm0.07$ in Figure~\ref{lae_fig:EndsleyPost}. This over-density is not as pronounced as the \citet{Hu2021} volume, and Poisson noise is larger because it has fewer than half the number of galaxies as the Hu volume. This results in a smaller inferred density and slightly larger error bars. While this region is quite likely to be an over-density, even this kind of association cannot (on its own) rule out a normal or slightly under-dense environment, though we have not factored the ionization environment in here.

%AT: Need to mention that this field was chosen because it's over-dense. That should affect the prior.

%%%%%%%%%%%%%%%%%%%%%%%%%%%%%%%%%%%%
\subsection{A JWST over-density}\label{lae_sec:JWST}

\begin{figure}
    \centering
    \includegraphics[width=0.485\textwidth]{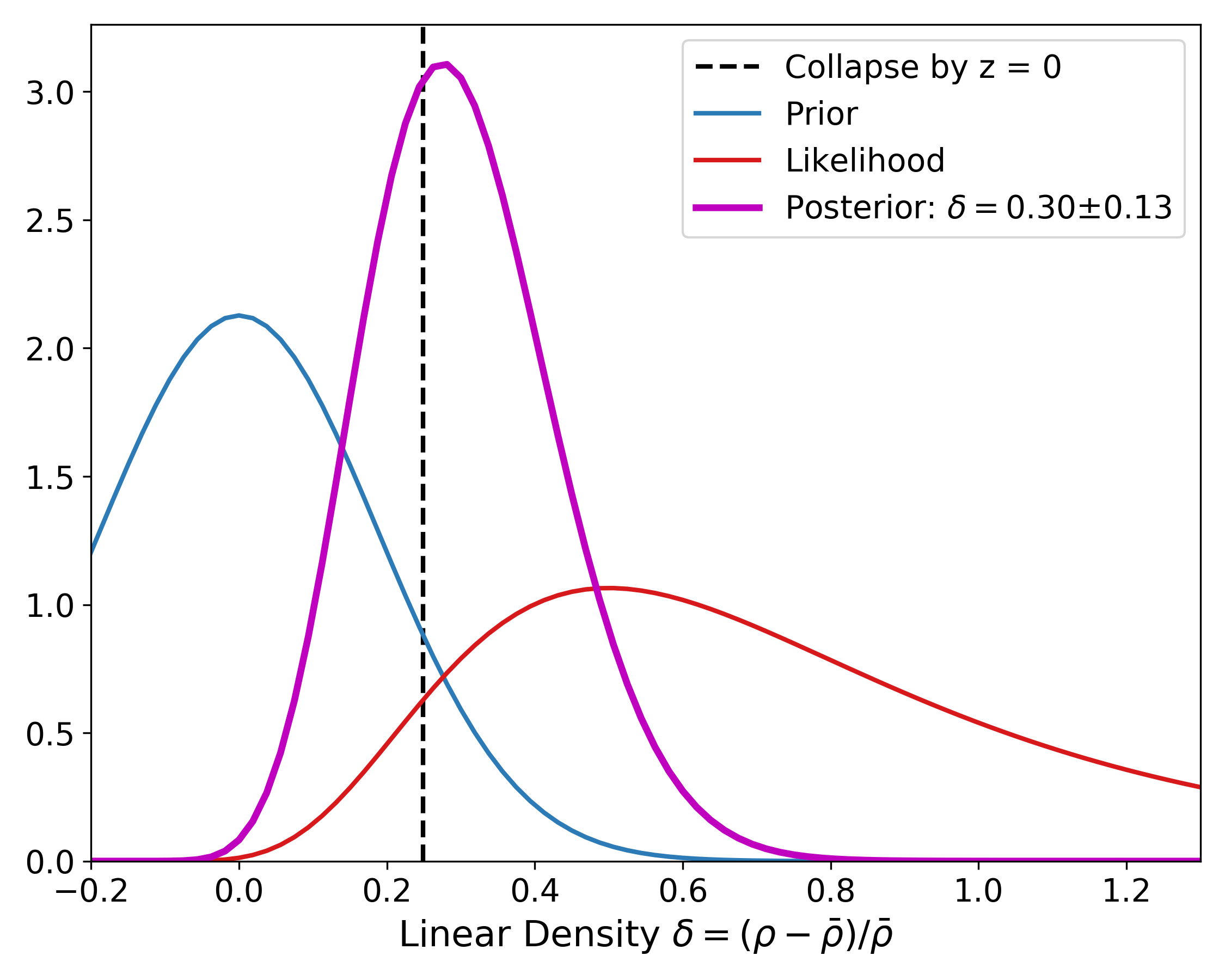}
    \caption{
    The inferred density  for the \citet{Laporte2022} volume, assuming no effect from reionization and that all eight observed galaxies are within $\Delta z=0.1$ of each other.  The \textit{magenta curve} shows posterior of the linear density, which has 68.27\% credible interval
    $\delta = 0.30 \pm 0.13$.
    The vertical dashed \textit{black} line indicates the approximate density required for this region to collapse into a cluster by $z = $ 0, $\delta_{\textrm{pc}} = 0.25$. The posterior (\textit{magenta}) indicates this volume is more than $50\%$ likely to be a protocluster, under our assumptions.
    The \textit{red and blue} curves show the likelihood and prior, respectively.}
    \label{lae_fig:JWSTPost}
\end{figure}

In \citet{Laporte2022}, an over-density of 8 galaxies (2 spectroscopically confirmed, 6 photometrically associated) are found in a $\sim20$ arcsecond radius region at $z \simeq 7.66$. \citet{Laporte2022} calculate the galaxy over-density of this region to be $\delta_\textrm{gal} = 4.0^{+2.4}_{-1.6}$, corresponding to an average number of sources expected in such a region to be $\left<N\right> = 1.6^{+0.75}_{-0.52}$. For this analysis, we assume the six photometrically-identified galaxies are within $\Delta z = 0.1$; the SED-fitting model uncertainties are actually much broader than this, so our analysis here is quite aggressive, essentially assuming that the galaxies will later be spectroscopically confirmed to be near each other. We estimate a bias value of $b = 6.4\pm1.1$ by using the \pakidge~package, assuming these galaxies are between $5 \times 10^9$ and $5 \times 10^{10}~M_\odot$; this is consistent with the $\sim 10^8~M_\odot$ stellar masses found by \citet{Laporte2022}. Using these values and assumptions, we measure the linear matter density to be $\delta = 0.30\pm0.13$ in Figure~\ref{lae_fig:JWSTPost}. The (\textit{purple}) posterior indicates that this volume is more than $50\%$ likely to be a virialize by the present day. The much smaller volume -- and thus higher cosmic variance in the dark matter density -- than the \citet{Endsley2022} or \citet{Hu2021} associations plays an important role in interpreting this association, making it much more likely to collapse in the first place. This can be seen in the density prior (\textit{cyan}), which has a significant portion of the pdf above the collapse threshold line. Given this over-density, we estimate the mass of the entire 20 arcsecond radius by $\Delta z = 0.1$ region to be $\sim 4 \times 10^{12} M_\odot$. The relatively small mass here indicates that collapse of the entire region by the present day is quite plausible; a search for a surrounding protocluster should be conducted on a larger angular scale. 

%%%%%%%%%%%%%%%%%%%%%%%%%%%%%%%%%%%%%%%%%%%%%%%%%%%%%%%%%%%%%%%%
%%%%%%%%%%%%%%%%%%%%%%%%%%%%%%%%%%%%%%%%%%%%%%%%%%%%%%%%%%%%%%%%
\section{Inference of the ionization field from LAE associations} 
%\section{In the context of a partially neutral Universe}
\label{lae_sec:methodspartial}

The last section assumed that \textit{no} LAEs 
%in the COSMOS or other LAGER fields 
were blocked by intervening neutral hydrogen, 
or in other words that the Universe was completely ionized (ionized fraction $\bar{Q} = 1$). In reality, at $z=$ 6.93, the Universe is thought to be only partially ionized (e.g., \citealt{Davies2018}).
In this case, we would expect only a fraction $\Fsurvavg$ of LAEs to actually be visible.
Thus, we would expect the true underlying $\Nexpavg$ to actually be \textit{larger} than the visible density of LAEs, $\Nexpavg = \Nvis / \Fsurvavg$ with $\Nvis = 1.5 \pm 0.1$ for the Hu volume\footnote{For the remainder of the paper, we will be using $\Nvis = 1.5 \pm 0.1$ calculated from all 4 LAGER fields in \citet{Wold2022} as our fiducial value.}. By itself, this would have the effect of shifting the density posterior of the \citet{Hu2021} volume to smaller over-densities. 

However, in this context a region with an over-density of visible LAEs must also be mostly ionized in order to see the sources inside it. In the standard picture of reionization (e.g., \citealt{Furlanetto2004}) it must have a \textit{high} density in order to be mostly ionized. More generally, in a fully ionized Universe, the clustering of LAEs is due to cosmic variance and Poisson noise alone.
In a mostly neutral Universe, ($\bar{Q} \ll 1$), we would expect to find nearly all of the visible LAEs (if indeed any could be found!) in a small number of large ionized regions -- or, in other words, more highly clustered. 
We found in section~\ref{lae_sec:resultsIonized} that the volume found in \citet{Hu2021} is highly clustered when compared to its surroundings, so that the probability of finding such a region among 20/118 same-sized regions in a fully ionized Universe is only 0.11\%/0.01\%. If ionized regions exaggerate the clustering, there must be a value for $\bar{Q} < 1$ that is most likely to produce a comparable LAE association exactly once within a larger survey volume.

Incorporating these considerations into the density posterior requires a mapping between between $\bar{Q}$ and $\Fsurvavg$ (sec.~\ref{lae_sec:Fsurvavg}) and a treatment of $\Fsurv$ for an individual region the size of the \citet{Hu2021} volume (sec.~\ref{lae_sec:Fsurv}).

%%%%%%%%%%%%%%%%%%%%%%%%%%%%%%%%%%%%%
\subsection{LAEs in a simple reionization model}
%\subsection{Fraction of LAEs that survive for whole Universe}
\label{lae_sec:Fsurvavg}

For a mapping between $\bar{Q}$ and $\Fsurvavg$, one first needs a model of reionization. In this section, we construct a very simple such model. Our prescription can be made more rigorous by comparing to more detailed reionization models, such as those generated by \texttt{21cmFAST} \citep{Mesinger2011, Murray2020,  Davies2021}, but we focus here on a simple prescription to make the inference framework as transparent as possible. 
This model was inspired by the measurement of a short mean free path  for ionizing photons,  $\lambda_i = 0.75^{+0.65}_{-0.45}$~pMpc at $z =6.0$ \citep{Becker2021}.

Let us assume that for an LAE to be visible, it must be inside an ionized bubble with $R > R_\alpha$ where $R_\alpha = 1$ pMpc, allowing for its ${\textrm{Ly}}\alpha$ photons to redshift out
of resonance \citep{MiraldaEscude1998}. Let us further assume that the ionizing photon mean free path is smaller than $R_\alpha$. This means that for an ionized bubble to grow large enough to allow transmission of Ly$\alpha$ photons, all of the ionizing photons that generate the bubble have to come from sources inside of it. Also, every region of size $R_\alpha$ is independent of its neighbors. In reality, some high-frequency photons will be shared between regions, but for the sake of simplicity we ignore them here. 

We make another simplifying assumption that the ionized fraction of hydrogen $Q$ in an independent region depends linearly on the fraction of baryons that have collapsed into haloes $f_{\textrm{coll}}$ via an efficiency parameter $\zeta$:
\begin{equation}\label{lae_eq:Qzetafcoll}
    Q = \zeta \cdot f_{\textrm{coll}}.
\end{equation}
This expression implicitly assumes that the ionizing efficiency of galaxies is independent of their mass. This is very unlikely to be true of real galaxies (e.g., \citealt{Trenti2010, Tacchella2013, Mason2015, Behroozi2015, Furlanetto2017}), but it allows for a very simple reionization model. In particular, within the Press-Schechter model \citep{Press1974, Lacey1993},
\begin{equation}\label{lae_eq:fcoll}
    f_{\textrm{coll}}(\delta, R_\alpha, z) = {\textrm{erfc}}\left( \frac{\delta_{\textrm{crit}}(z) - \delta_0}{\sqrt{2(\sigma^2_{\textrm{min}} - \sigma^2_{R_\alpha})}} \right),
\end{equation}
where $\delta_{\textrm{crit}}(z)$ is the linearized density threshold for spherical collapse \citep[approximately 1.69 divided by the growth factor of dark matter structure,][]{Eisenstein1998}, $\delta_0$ is the linearized density of the region $\delta$ scaled to $z = 0$ (again, via the growth factor), $\sigma_{R_\alpha}$ is the linear r.m.s fluctuation of the dark matter density field on the scale of $R_\alpha$, and $\sigma_{\textrm{min}}$ is the same on the scale of the smallest virialized halo allowed to form a galaxy. We take that smallest scale to correspond to a halo virial temperature $T_{vir}=10^4K$, when atomic line cooling becomes efficient enough for gas clouds to collapse and fragment for star formation \citep{Loeb2013}.

Note that we assume that the ionized fraction precisely follows the underlying density of the field -- and thus we ignore Poisson fluctuations in the galaxy counts! This is of course inconsistent with our inference model, but we note that the observed galaxy population is just the tip of the iceberg -- if it is 2.5$\sigma$ over-dense, the Hu volume is expected to have $\sim$40,000 galaxies above $M_\textrm{abs} = -13$, and $\sim$3 galaxies above the characteristic luminosity $M_\textrm{abs} = -21.15$ (using the UV luminosity function from \citealt{Bouwens2021} and cosmic variance model from \citealt{Trapp2020}). Thus Poisson fluctuations for the entire galaxy population are far smaller than for the small fraction observable as LAEs in existing surveys. 

\begin{figure}
    \centering
    \includegraphics[width=0.485\textwidth]{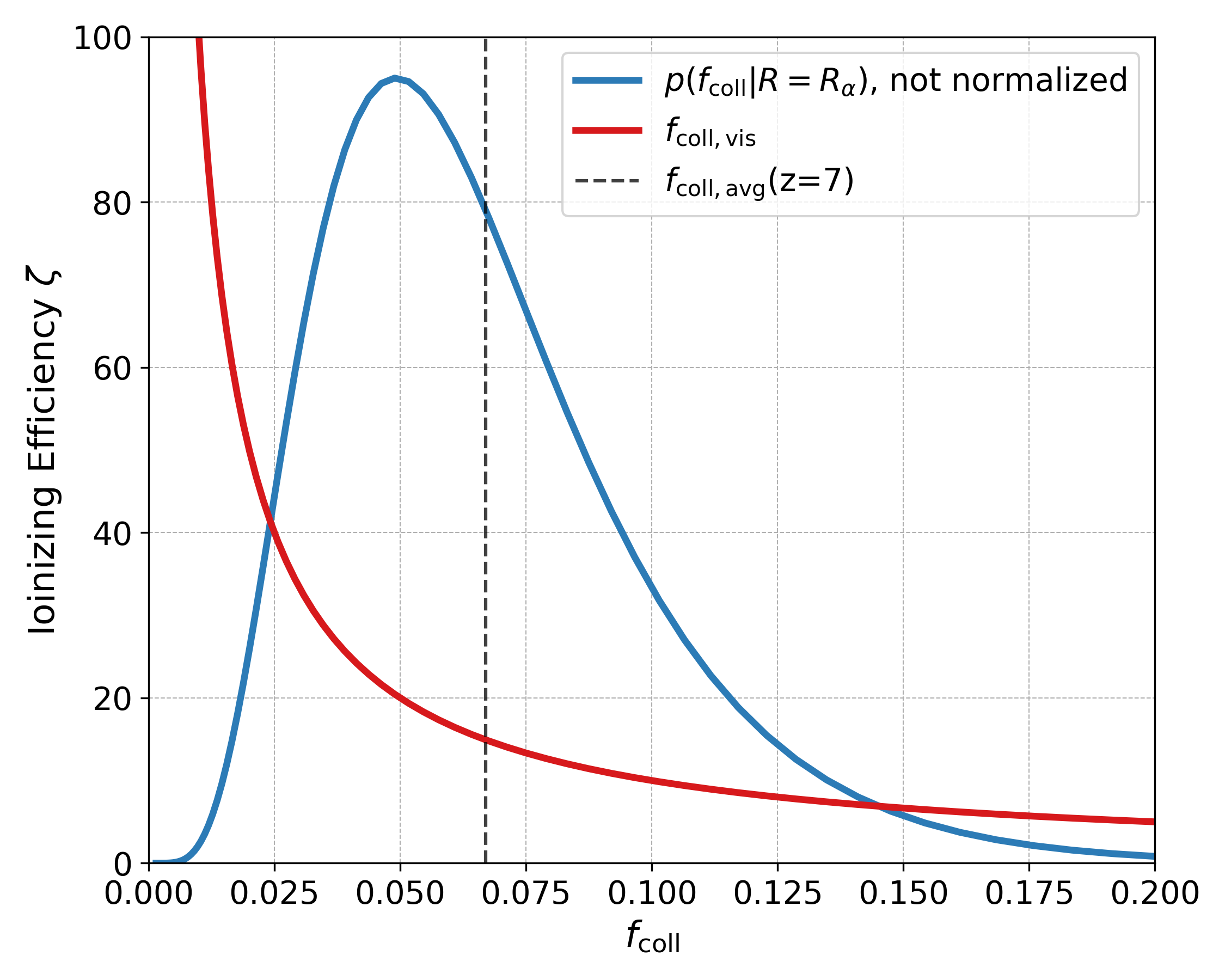}
    \caption{
    Properties of our simple reionization model. The 
    \textit{red} curve shows the ionizing efficiency required to ionize an independent region as a function of its collapse fraction (see eq.~\ref{lae_eq:Qzetafcoll}).
    The \textit{blue} curve shows $p(f_{\textrm{coll}}|R=R_\alpha)$, the (un-normalized) probability distribution of collapse fractions for regions of size $R = R_\alpha$. The \textit{black dashed} curve shows the  average collapse fraction at $z=$ 7.}
    \label{lae_fig:zetafcoll}
\end{figure}

In this simple model in which the ionized fraction increases monotonically with the local collapse fraction
$f_{\textrm{coll}}$, a region is fully ionized if it has a sufficiently high  density. Setting $Q = 1$, we obtain a relationship between the efficiency $\zeta$ and $f_{\textrm{coll,vis}}$, the collapse fraction required to ionize a region of size $R_\alpha$: $\zeta = 1/f_{\textrm{coll,vis}}$ (see Fig.~\ref{lae_fig:zetafcoll}, \textit{red} curve). From equation~(\ref{lae_eq:fcoll}), we can also obtain $p(f_{\textrm{coll}}|z, R)$ (see Fig.~\ref{lae_fig:zetafcoll}, \textit{blue} probability distribution).

Choosing a value for $f_{\textrm{coll,vis}}$ then defines the fraction of equal-mass regions at each redshift that can contain visible LAEs:
    \begin{equation}
        \bar{F}_{\textrm{vis}}(z) = \int_{f_{\textrm{coll,vis}}}^{1} p(f_{\textrm{coll}}|z, R_\alpha) df_{\textrm{coll}}.
    \end{equation}
Weighting by the number of LAEs inside a region of $f_{\textrm{coll}}$ gives the fraction of LAEs in that are visible at each redshift:
\begin{equation}
    \Fsurvavg(z) = \frac{1}{\Nexpavg} \int_{f_{\textrm{coll,vis}}}^{1} N_{\textrm{exp}}(f_{\textrm{coll}}) \cdot p(f_{\textrm{coll}}|z, R_\alpha) df_{\textrm{coll}},
\end{equation}
where $\Nexpavg$ is the average number of LAEs in a region of size $R_\alpha$ and $N_{\textrm{exp}}(f_{\textrm{coll}})$ is the number of LAEs in a region of size $R_\alpha$ but with collapse fraction $f_{\textrm{coll}}$. Since each collapse fraction has a unique $\delta$ value on this scale, $N_{\textrm{exp}}(f_{\textrm{coll}}(\delta)) = \Nexpavg (1+\delta \cdot b)$ (like eq.~\ref{lae_eq:conditional1}).
Similarly, we can weight by the ionization of each region through $Q = f_{\textrm{coll}} / f_{\textrm{coll,vis}}$ (taking $Q = 1$ for $f_{\textrm{coll}} > f_{\textrm{coll,vis}}$) to obtain the average ionization of the Universe at some redshift:
    \begin{equation}
        \bar{Q}(z) = \int_{0}^{1} \frac{f_{\textrm{coll}}}{f_{\textrm{coll,vis}}} \cdot p(f_{\textrm{coll}}|z, R_\alpha) df_{\textrm{coll}}.
    \end{equation}
Note that $\bar{Q}$ is \emph{not} simply the globally-averaged collapse fraction multiplied by $\zeta$, because some photons in over-dense regions are wasted thanks to absorption by small-scale features: this is important because we have assumed $\lambda_i$ to be smaller than the size of the independent regions.

\begin{figure}
    \centering
    \includegraphics[width=0.485\textwidth]{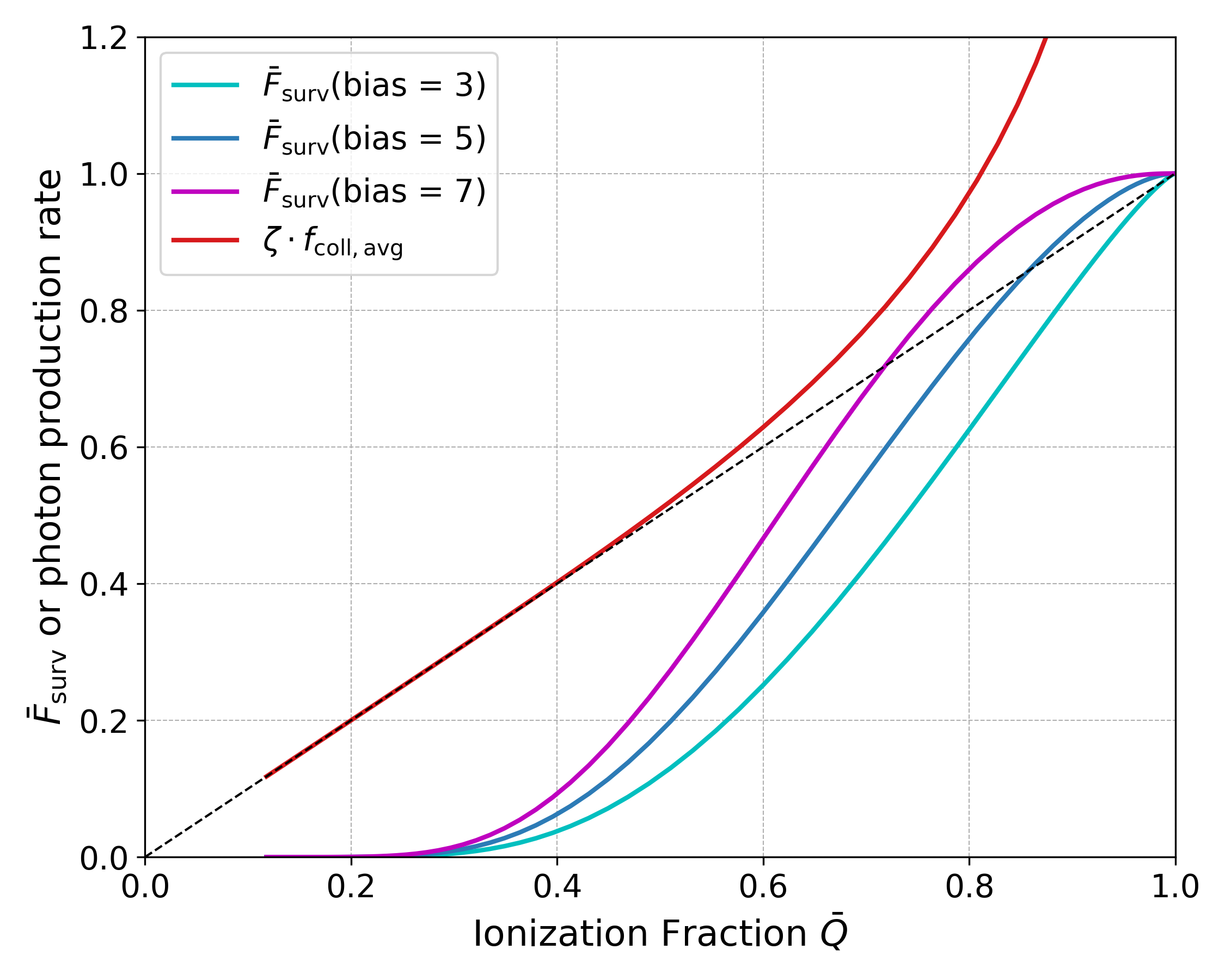}
    \caption{
    The fraction of LAEs that are visible $\Fsurvavg$ as a function of average ionization fraction $\bar{Q}$ for various bias values.
    Higher bias values result in a larger fraction of all LAEs being visible for a given ionization fraction of the Universe.
    The \textit{red} line shows the total ionizing photon production rate $\zeta \cdot f_{\textrm{coll,avg}}$, for which a value above the line $x=y$ means photons are escaping their host galaxies but being absorbed before contributing to reionization. The most over-dense $R_\alpha$ regions are overproducing photons but can't help their neighbors reionize. This effect is only large when the Universe is mostly reionized.}
    \label{lae_fig:Fsurvive}
\end{figure}

Finally, we obtain a relationship between the average ionized fraction of the universe $\bar{Q}(z)$ and the fraction of LAEs that are visible $\Fsurvavg(z)$ (averaged across the Universe) for any choice of $f_{\textrm{coll,vis}}$ (as shown in Fig.~\ref{lae_fig:Fsurvive}). 
Higher bias values result in a larger fraction of all LAEs being visible for a given ionization fraction of the Universe. We also plot in Figure~\ref{lae_fig:Fsurvive} the total ionizing photon production rate relative to the number of hydrogen atoms, $\zeta \cdot f_{\textrm{coll,avg}}$. A value above the line $x=y$ implicitly means photons are escaping their host galaxies, but being absorbed before contributing to reionization. The most over-dense $R_\alpha$ regions are overproducing photons but can't help their neighbors reionize. This effect is only significant once the Universe is mostly reionized.

We have made another important simplification here by assuming that the survival fraction of LAEs is determined entirely by the local ionization environment, which is itself determined by the \emph{average} collapse fraction given by the underlying density. LAEs in partially ionized volumes that happen to have ionized regions along the line of sight will survive, while LAEs in fully ionized regions may not if they sit near mostly neutral regions. Moreover, the amount of damping wing absorption is determined by a broad path length through the IGM, so its effects depend on inhomogeneous reionization  \citep{MiraldaEscude1998, Mesinger2008b}. 

Additionally, we have not incorporated Poisson fluctuations in the galaxy counts into the \emph{local} $\Fsurv$ calculation. That is, even if a volume has an over-density of galaxies due to an upward Poisson fluctuation rather than a true matter over-density, it will still be ``over-ionized" relative to the rest of the Universe. This will broaden the distribution of $\Fsurv$ and hence slightly weaken our constraints.

Finally, we also note that our model is similar in spirit to \citet{Furlanetto2004}, which builds the ionization field from the density field but assuming that the ionizing photons have long mean free paths. That model provides a useful qualitative picture of reionization but systematically underestimates the bubble sizes in comparison to semi-numeric simulations that use the same source models \citep{Lin2016}. This would tend to moderate the modulation induced by reionization on the LAE population.

%%%%%%%%%%%%%%%%%%%%%%%%%%%%%%%%%%%%%
\subsection{LAE galaxy associations in a partially ionized Universe}\label{lae_sec:Fsurv}
%\subsection{LAE galaxy associations in an ionized Universe} 
%\subsection{Fraction of LAEs that survive within a finite region of given density}

In the previous subsection, we examined the visibility of LAEs in regions of size $R_\alpha$, the minimum ionized bubble size to host such sources. But in practice, galaxy associations may subtend significantly larger scales in real surveys.
The \citet{Hu2021} volume is larger than $R_\alpha$, meaning it may have some sub-chunks that are ionized, and some that are not. In this section, we consider the distribution of $\Fsurv(\delta,R > R_\alpha)$: the fraction of LAEs that are visible in a region of size $R > R_\alpha$ and density $\delta$.

Such a region has $N = (R/R_\alpha)^3$ sub-chunks of size $R_\alpha$, each with a density $\delta_i$ distributed around $\delta$ with standard deviation $\sigma_{\textrm{w}} = \sqrt{\sigma^2_{R_\alpha} - \sigma^2_R}$, where $\sigma_{R_\alpha}$ and $\sigma_R$ are the rms density fluctuation of the dark matter on a scale of $R_\alpha$ and $R$, respectively. Each chunk then also has corresponding $f_{\textrm{coll,i}}$.
Each of those sub-chunks are either ionized ($f_{\textrm{coll,i}} > f_{\textrm{coll,vis}}$, allowing LAEs to be visible) or not (so that their LAEs are invisible).
We can then calculate $F_{\textrm{vis}}$, $\Fsurv$, and $Q$ for the region of size $R$ via weighted averages over the sub-chunks rather than integrals (as we did in the last section). Note that we ignore correlations between the sub-regions. 

\begin{figure*}
    \centering
    \includegraphics[width=1.0\textwidth]{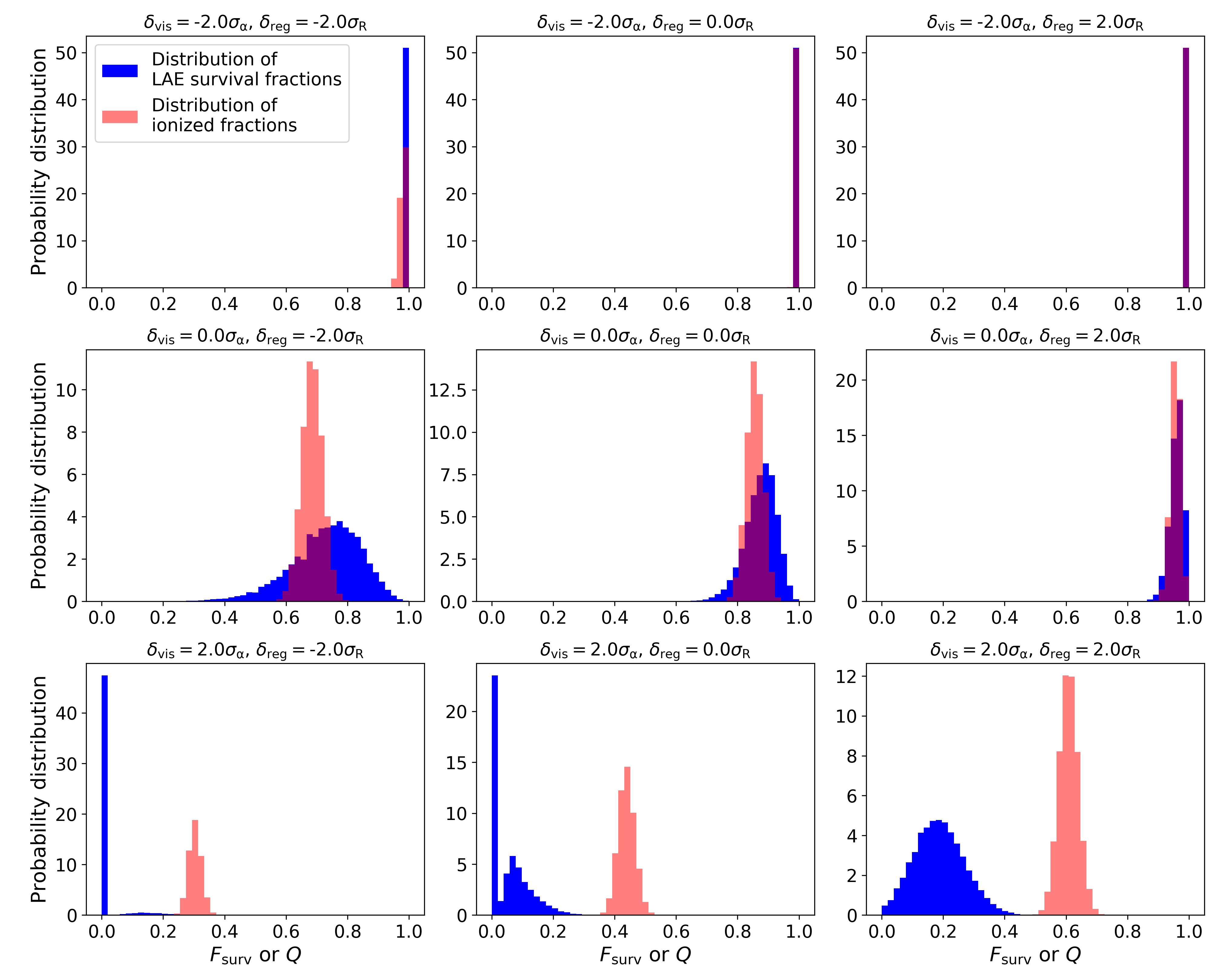}
    \caption{The distribution of the LAE survival fraction $\Fsurv$ (blue histograms) and local ionized fraction (red histograms) for regions 
    matching the \citet{Hu2021} volume.
    The panels vary the density $\delta_{\textrm{reg}}$ of that region (left to right) and the density required for a region of size $R_\alpha$ to fully ionize, $\delta_{\textrm{vis}}$ (top to bottom), 
    or equivalently the ionizing efficiency of dark matter haloes.
    $\Fsurv$ can vary widely between regions of the same size and density (more so than $Q$ in those regions!), due to the effects of Poisson noise. When the value of $\delta_{\textrm{vis}}$ is large ($\bar{Q} \ll 1$), the distribution of $\Fsurv$ can become bimodal, with a large fraction of regions having zero visible LAEs. Here we set $b= 7.31$.
    }
    \label{lae_fig:FsurvDist}
\end{figure*}

In Figure~\ref{lae_fig:FsurvDist}, we show how $\Fsurv$ can vary widely between volumes of the same size $R$ and overall density $\delta$ (more so than $Q$ in those regions!), due to the effects of Poisson noise and density fluctuations on the sub-chunk scale $R_\alpha$.
When the value of $f_{\textrm{coll,vis}}$ is large (or equivalently when $\delta_{\textrm{vis}}$ is large and $\bar{Q} \ll 1$), the distribution of $\Fsurv$ can become bimodal, with a large fraction of regions having zero visible LAEs and all of the visible LAEs confined to a few very over-dense regions.

%%%%%%%%%%%%%%%%%%%%%%%%%%%%%%%%%%%%%
\subsection{The inference framework}
%\subsection{Monte-Carlo generation of Likelihood}
\label{lae_sec:MC}

Within the framework described above, we now imagine that a survey has found an association of LAEs, and we use the model to measure three interesting quantities: \emph{(a)} the local density of the region (as in section~\ref{lae_sec:resultsIonized}), \emph{(b)} the local ionized fraction in the region, and \emph{(c)} the \emph{global} ionized fraction. We next describe how we make such inferences in practice. This procedure must be tuned to the specific construction of the survey; here, we imagine the simple case of choosing the most extreme apparent over-density in a survey, similar to the method of \citet{Hu2021}. 
(One difference is that our method implicitly assumes the region volumes and tiling strategy are chosen \emph{before} the survey is complete rather than chosen ``by eye" afterward -- a process that is difficult to quantify statistically.)

We generate the likelihood of finding $N_{\textrm{LAE}}$ sources in a region with \textit{(i)} radius $R > R_\alpha$ pMpc among a larger survey with $N_{\textrm{vols}}$ regions of the same size, \textit{(ii)} average visible LAE number  $\Nvis$, \textit{(iii)} density $\delta$, and \textit{(iv)} collapse fraction required for sub-chunks to be ionized $f_{\textrm{coll,vis}}$ by running a Monte-Carlo simulation.

\begin{enumerate}
    \item First, we choose a value for $f_{\textrm{coll,vis}}$, which defines $\Fsurvavg$ and $\bar{Q}$. As shown in Figure~\ref{lae_fig:zetafcoll}, this implicitly determines the effective ionizing efficiency $\zeta$. This also defines the \textit{actual} expected number of LAEs through $\Nexpavg = \Nvis / \Fsurvavg$.
    \item Next, we generate a set of $N_{\textrm{vols}}$ volumes each with a dark matter density $\delta$ drawn from a normal distribution with standard deviation $\sigma_R$. Each of those volumes has $N_{\textrm{sub}} = (R/R_\alpha)^3$ sub-chunks with densities $\delta_{\textrm{sub}}$ drawn from a normal distribution\footnote{When drawing this way, the sub-chunk densities $\delta_{\textrm{sub}}$ will not add up to exactly $\delta$. However, with the large number of sub-chunks considered in this paper, the deviation is small. When $N_{\textrm{sub}}$ is small, $\delta$ must be re-calculated from the average of all $\delta_{\textrm{sub}}$ values. However, this process imposes a broadening in the distribution of $\delta$ that must be corrected for.} centered at $\delta$ with standard deviation $\sigma_{\textrm{w}}$. These densities each have corresponding $f_{\textrm{coll,sub}}$.
    \item The number of sources expected per sub-chunk is $N_{\textrm{exp,sub}} = N_{\textrm{avg, sub}} (1 + b \cdot \delta_{\textrm{sub}})$, with $N_{\textrm{avg, sub}} = \Nexpavg / N_{\textrm{sub}}$. We then draw from a Poisson distribution for each subchunk with $\lambda = N_{\textrm{exp,sub}}$ to get the number of LAEs per subchunk $N_{\textrm{draw,sub}}$. Finally, we sum all LAEs in subchunks with $f_{\textrm{coll,sub}} > f_{\textrm{coll,vis}}$ to get the total number of \textit{observable} LAEs. Then, out of the $N_{\textrm{vols}}$ volumes, we keep the one with the most observed LAEs, mimicking the procedure of \citet{Hu2021}.
    \item We repeat (i) - (iii) many times, each time choosing a random value for $\Nvis$ and $b$ according to their uncertainty. The likelihood $P(N_{\textrm{LAE}}|b,f_{\textrm{coll,vis}})$ is the fraction of volumes chosen in step (iii) that have the correct number of observed LAEs ($N_{\textrm{obs}} = N_{\textrm{LAE}}$) 
    %\textit{and} are mostly ionized ($\Fsurv > 0.5$, see below for an explanation of this cutoff).
    This step implicitly marginalizes over the density $\delta$ and uncertainty in $\Nvis$ and $b$.
    \item\label{test} We then repeat (i)-(iv) for many values of $f_{\textrm{coll,vis}}$, each having a corresponding $\Fsurvavg$ and $\bar{Q}$ (depending on the random draw of $b$). This allows us to construct the likelihood as a function of $\bar{Q}$.
    \item In order to transform the likelihood into a posterior distribution, we require a prior on the intrinsic LAE density. Many such choices are possible; we multiply by a prior that enforces that the number density of LAE hosts (whether visible as LAEs or not) must \textit{increase} monotonically as redshift decreases (see below for an explanation of this prior).
    \item Finally, we take all volumes selected above across every value of $f_{\textrm{coll,vis}}$, weight by our prior, and make a histogram of the densities $\delta$. This histogram is the posterior of $\delta$ marginalized over $\bar{Q}$. Similarly, we take these volumes, weight by the prior, and construct the posterior around the ionized fraction of the LAE association, marginalized over $\bar{Q}$ and $\delta$.
\end{enumerate}

%%%%%%%%%%%%%%%%%%%%%%%%%%%%%%%%%%%%%
\subsection{The choice of priors} \label{lae_sec:priors}

The most important prior in our framework is that on the underlying density of a region, which is, to linear order, simply a gaussian centered at zero with standard deviation $\sigma_R$. 
This prior is determined by integrating the power-spectrum of dark matter fluctuations over the desired physical scale \citep[see e.g.,][]{Newman2002, Stark2007, Munoz2010, Robertson2010, Trapp2020} to obtain $\sigma_R$. This prior is highly constraining; in sections~\ref{lae_sec:resultsIonized} and~\ref{lae_sec:resultsPartiallyIonized}, the posterior of the density is prior-dominated.
Fortunately, this prior is also very well-specified by a variety of cosmological probes. 

A second prior, limiting the underlying galaxy density, also turns out to be important. In practice, one reasonable solution for a survey with a single, large LAE association and no other visible LAEs would be a highly neutral Universe with a single large ionized bubble. However, this would require that the true number density of LAE hosts be much larger than the observed density, because most of the Lyman-$\alpha$ lines are attenuated by the IGM. It is thus helpful to include a prior on the underlying density of the host galaxies. There are a variety of ways one can do this; we make a conservative choice here. 

In particular, we include a prior requiring that there are not intrinsically more galaxies capable of hosting LAEs -- for example -- at $z =$ 6.93 than at $z =$ 5.7. In reality, there are likely many more LAEs at $z =$ 5.7 than $z =$ 6.93, 
because the underlying luminosity function of galaxies is evolving rapidly, so this serves as a conservative bound. \citet{Wold2022} finds the LAE luminosity density ratio between $z = $ 6.9 and 5.7 to be $\rho_{z=6.9} / \rho_{z=5.7} = 0.63^{+0.13}_{-0.15}$. 
As long as the number density of LAEs stayed constant or increased from $z = $ 6.9 to 5.7
(and approximating the mean LAE luminosity as a constant), this measurement can be converted to a lower-bound on the value of $\Fsurvavg(z = 6.9) > 0.63_{-0.15}$. We then convert to a lower bound on $\bar{Q}$ using our mapping between $\Fsurvavg$ and $\bar{Q}$ (see Fig.~\ref{lae_fig:Fsurvive}).
Again, this is a conservative approach, because there are many fewer galaxies overall at $z\sim 7$ than at $z \sim 6$. One could incorporate a full model for the LAE population as a more sophisticated prior, 
or else simultaneously constrain both the overall galaxy evolution and the LAE distribution.

Finally, we use the bias value and its uncertainty from \citet{Hu2021} \citep[in turn taken from][]{Ouchi2018} of $b = 7.31 \pm1.01$ (after converting to a linear bias). A theoretical model of cosmic variance predicts a value of the linear bias between 2-7 for halo masses between 10$^9$ and 10$^{12} M_\odot$
at this redshift and scale \citep{Trapp2020}, 
implying these LAEs are on the massive end. However, a full comparison between observationally-obtained bias values and theoretical values is outside the scope of this paper.
Further,
we have shown in section~\ref{lae_sec:resultsIonized} and will show again in section~\ref{lae_sec:resultsPartiallyIonized} that the bias value has surprisingly little effect on the posteriors, likely owing to the fact that the prior on the linear density is dominant.

%%%%%%%%%%%%%%%%%%%%%%%%%%%%%%%%%%%%%%

%%%%%%%%%%%%%%%%%%%%%%%%%%%%%%%%%%%%%%%%%%%%%%%%%%%%%%%%%%%%%%%%
%%%%%%%%%%%%%%%%%%%%%%%%%%%%%%%%%%%%%%%%%%%%%%%%%%%%%%%%%%%%%%%%

%%%%%%%%%%%%%%%%%%%%%%%%%%%%%%%%%%%%%%%%%%%%%%%%%%%%%%%%%%%%%%%%
%%%%%%%%%%%%%%%%%%%%%%%%%%%%%%%%%%%%%%%%%%%%%%%%%%%%%%%%%%%%%%%%

\section{Applying the full inference framework}
%\section{Results for partially ionized Universe}
\label{lae_sec:resultsPartiallyIonized}

\begin{figure}
    \centering
    \includegraphics[width=0.485\textwidth]{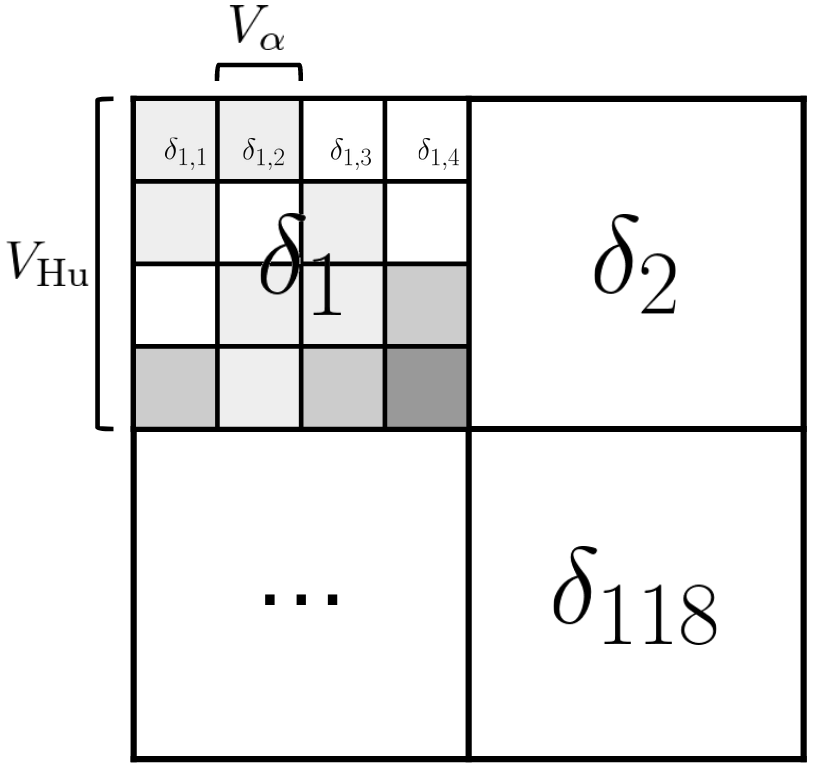}
    \caption{A simulation of the four volumes described in \citet{Hu2021} and \citet{Wold2022}, split into 118 sub-volumes, each the size of the \citet{Hu2021} over-density ($V_\textrm{Hu}$). Each of those volumes are assigned a density $\delta_i$ and then split into sub-chunks (51 each, volume $V_\alpha$) each with a density $\delta_{i,j}$ drawn from around the density $\delta_i$. The $V_\alpha$ cells are shaded to represent different ionization environments; only those that are fully ionized would contain visible LAEs. See Section~\ref{lae_sec:MC} for a description of how many such simulations are used to generate a likelihood and posterior.}
    \label{lae_fig:CartoonMC}
\end{figure}

As an example of the inference framework, we now apply our procedure to the LAE associations observed by \citet{Hu2021} and \citet{Endsley2022}, focusing on the former as the more extreme case. For the Hu volume, $R = 3.7$ pMpc giving $N_{\textrm{sub}} = 51$, $N_{\textrm{vols}} = 118$ representing the total volume of all LAGER fields \citep{Wold2022}, $\Nvis = 1.5 \pm 0.1$, and $b = 7.31 \pm 1.02$ (see section~\ref{lae_sec:resultsIonized}).
Figure~\ref{lae_fig:CartoonMC} shows a diagram of the MC simulation layout for this case, with $V_\textrm{Hu}$ and $V_\alpha$ corresponding to $R = 3.7$ pMpc and $R_\alpha$ (see sec.~\ref{lae_sec:methodspartial}).

%%%%%%%%%%%%%%%%%%%%%%%%%%%%%%%%%%%%%
\subsection{What can we learn about reionization?}

\begin{figure}
    \centering
    \includegraphics[width=0.485\textwidth]{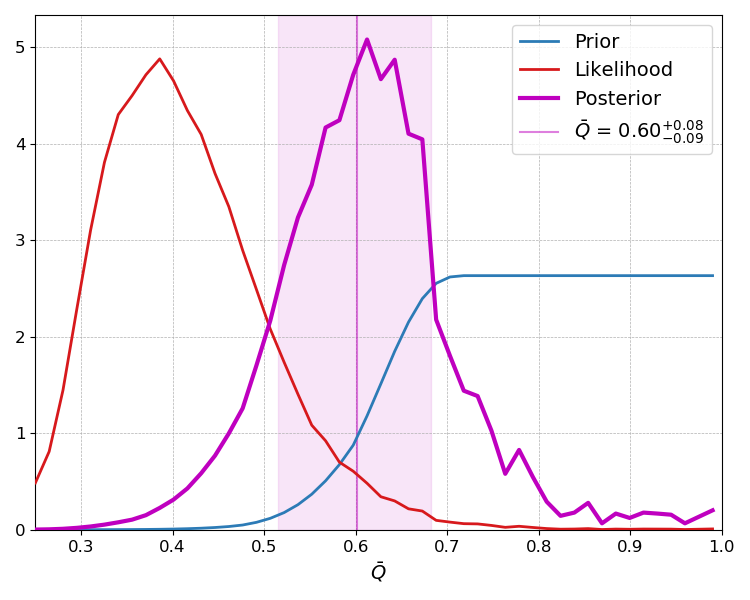}
    \caption{Constraints on the global ionized fraction $\bar{Q}$ from the \citet{Hu2021} observation. The \textit{magenta} curve shows the posterior on the ionization of the Universe and 68.27\% credible interval $\bar{Q} = 0.60^{+0.08}_{-0.09}$ (statistical errors only). 
    The \textit{red} curve shows the likelihood of finding 14 sources given the average ionization of the Universe $\bar{Q}$ at $z =$ 6.93 and other model parameters, while the 
    \textit{blue} curve shows the prior on the ionized fraction of the Universe, assuming the intrinsic number of LAEs increased from $z=$ 6.93 to 5.7.
    At $\bar{Q} \lesssim 0.6$, the posterior is prior-dominated.
    }
    \label{lae_fig:QbarPost}
\end{figure}

Figure~\ref{lae_fig:QbarPost} shows the likelihood and posterior of the globally-averaged ionized fraction of the Universe at $z =$ 6.93. The median and inner 68.27\% probability of the posterior is $\bar{Q} = 0.60^{+0.08}_{-0.09}$. 
% The posterior is perhaps better represented as a 2$\sigma$ lower limit on the ionization, $\bar{Q} > 0.55$. 
%1-sigma lower limit is $\bar{Q} > 0.65$
At $\bar{Q} \lesssim 0.6$, the posterior is prior-dominated. The prior represents a case where the number density of LAEs \textit{stayed the same} between $z=$6.9 and 5.7. In reality, the number density likely increased dramatically, which would push the posterior to even higher values of $\bar{Q}$.

\begin{figure}
    \centering
    \includegraphics[width=0.485\textwidth]{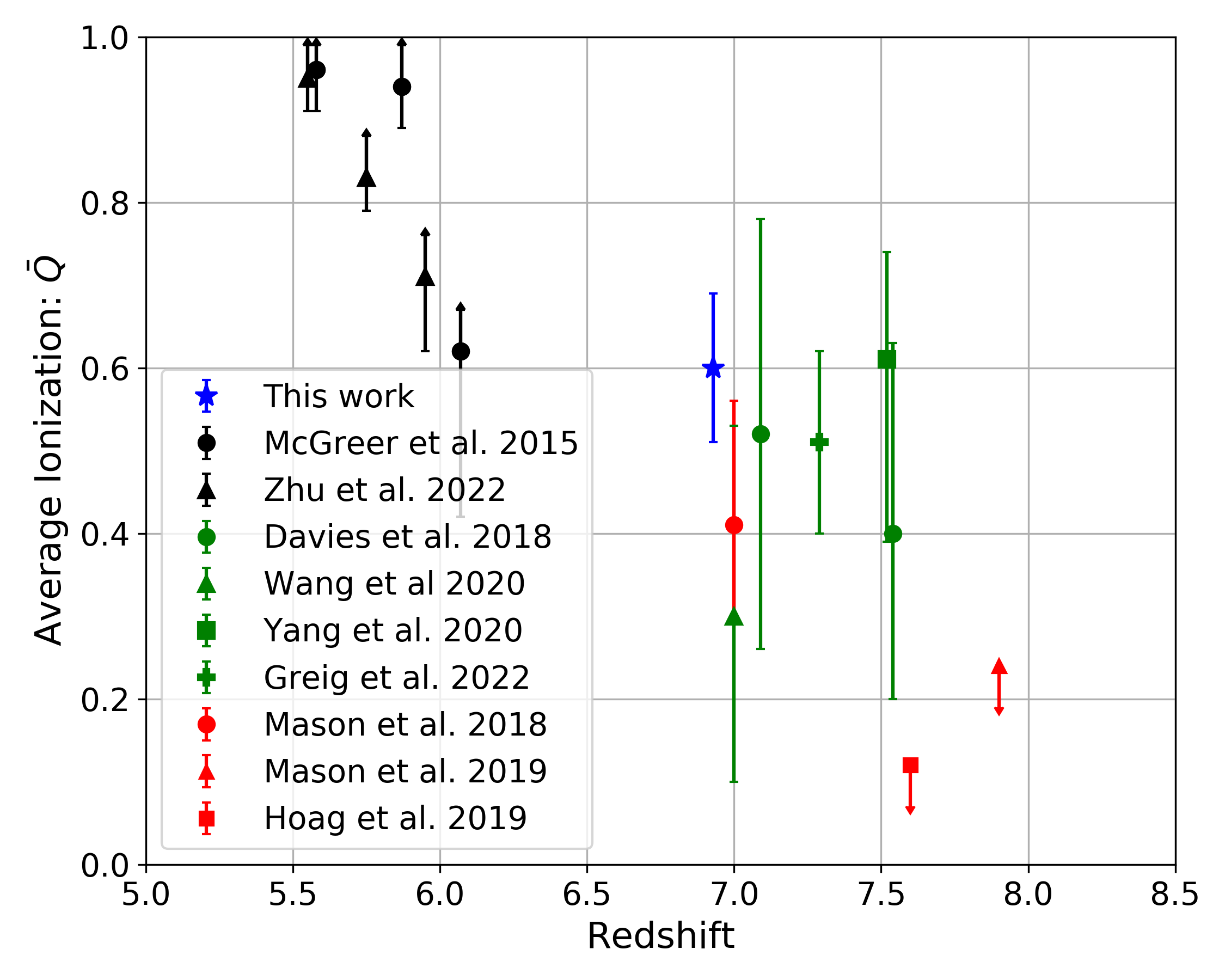}
    \caption{Comparison of our results with others in the literature. In \textit{black} are methods using QSO dark gaps; in \textit{green} are methods using QSO damping wings; and in \textit{red} are methods using Lya emission of LBGs. Our simple model (\textit{blue}) agrees with other measurements that suggest reionization is incomplete (but relatively advanced) at $z \sim 7$. Note that the true error bars on our measurement are larger, because of systematic uncertainties with our model of reionization. }
    \label{lae_fig:QbarCompare}
\end{figure}

We note that our analysis disfavors both small ionized fractions (largely due to the prior on the LAE abundance) and a nearly ionized Universe. The latter is perhaps the most interesting aspect, as the constraint comes from finding a single region with so many LAEs and is not driven by any of our priors, although the quantitative constraint does depend upon our reionization prescription and can be improved with more detailed models of that process.  Thus our simple model agrees with other measurements that suggest reionization is incomplete (but relatively advanced) at $z \sim 7$
(see Fig.~\ref{lae_fig:QbarCompare}, \citealt{McGreer2015,Mason2018,Davies2018,Mason2019,Hoag2019,Wang2020,Yang2020,Greig2022,Zhu2022}). %Greig2017,Inoue2018,

%%%%%%%%%%%%%%%%%%%%%%%%%%%%%%%%%%%%%
\subsection{Is the association inside a large ionized bubble?}

\begin{figure}
    \centering
    \includegraphics[width=0.485\textwidth]{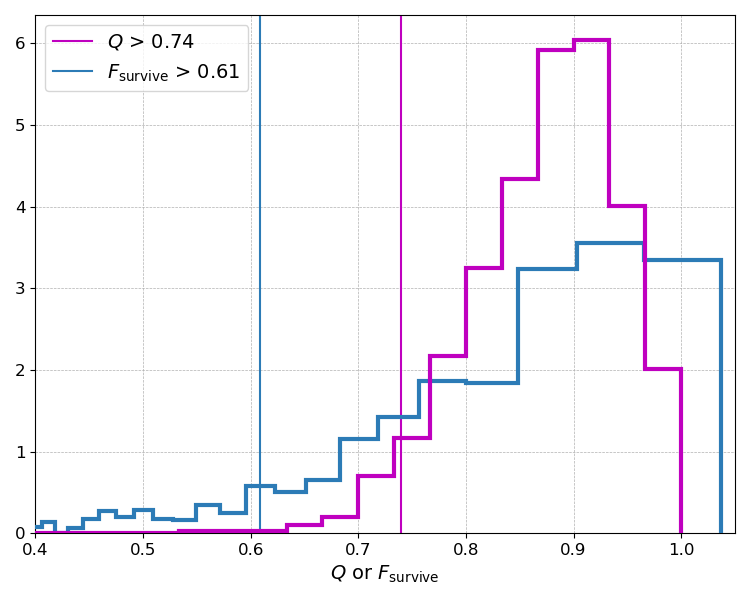}
    \caption{The \citet{Hu2021} volume is highly ionized with most bright LAE hosts visible as line emitters, as suspected by those authors. The \textit{magenta} histogram shows the ionization states $Q$ from each MC simulation that resulted in finding 14 sources, weighted by the prior on $\bar{Q}$. The \textit{blue} curves the same but for $\Fsurv$, the fraction of LAEs that are visible. The vertical bars indicate the 95.45\% lower limits of these quantities, $Q > 0.74$, $\Fsurv > 0.61$.
    }
    \label{lae_fig:HuQsFsurvives}
\end{figure}

Figure~\ref{lae_fig:HuQsFsurvives} shows the histogram of the local ionization states $Q$ from
each MC simulation that resulted in finding 14 sources, weighted
by the prior on $\bar{Q}$, and the same for $\Fsurv$, the fraction of LAEs in the \citet{Hu2021} volume that are visible. The 95.45\% credibility lower limits for these quantities are $Q > 0.74$ and $\Fsurv > 0.61$. 
Thus the qualitative mapping of this LAE association to a large ionized region is largely validated: the volume is highly ionized with high credibility. Moreover, the fraction of LAE hosts that are visible as line emitters is also quite large.

%%%%%%%%%%%%%%%%%%%%%%%%%%%%%%%%%%%%%
\subsection{Is the association a protocluster?} 

\begin{figure}
    \centering
    \includegraphics[width=0.485\textwidth]{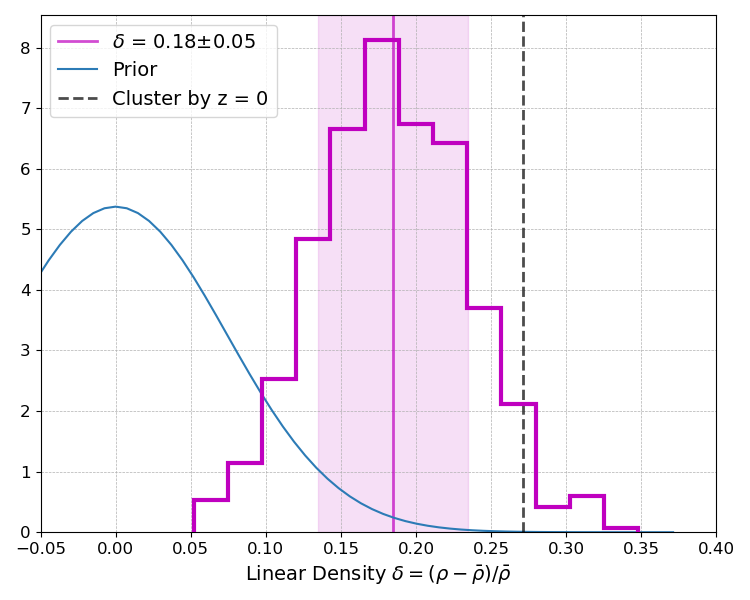}
    \caption{Constraints on the overall density of the \citet{Hu2021} volume. The \textit{magenta} curve shows the histogram of all densities that resulted in finding 14 sources in the MC simulation described in section~\ref{lae_sec:MC}, weighted by the prior on $\bar{Q}$.
    This gives a 68.27\% credible interval of $\delta = 0.18 \pm 0.05$, only slightly lower than the results from the fully-ionized case (see Fig.~\ref{lae_fig:HuPost_COSMOS}). The entire region is not likely to have collapsed into a single cluster by $z =$ 0, which would require a linear density of $\delta_{\textrm{pc}} = 0.27$ (\textit{black dashed} line). 
    The prior on the density is shown in \textit{blue}.
    }
    \label{lae_fig:DenPost}
\end{figure}

Figure~\ref{lae_fig:DenPost} shows the posterior of the density, which yields a 68.27\% credible interval of $\delta = 0.18\pm0.05$, only slightly lower than the results from the fully-ionized case (see Fig.~\ref{lae_fig:HuPost_COSMOS}). Given the r.m.s. density fluctuation in the linear density field for the \citet{Hu2021} volume $\sigma_R=$0.074, this corresponds to a $2.4 \pm 0.7$ $\sigma_R$ over-density.
Despite its extreme apparent over-density, the entire region is not likely to have collapsed into a single cluster by $z =$ 0, which would require a linear density of $\delta_{\textrm{pc}} = 0.27$. However, it is possible that one or more sub-regions within this larger volume are at sufficiently high density to collapse into clusters by $z =$ 0.

%%%%%%%%%%%%%%%%%%%%%%%%%%%%%%%%%%%%%
\subsection{Sensitivity to model parameters}

\begin{figure}
    \centering
    \includegraphics[width=0.485\textwidth]{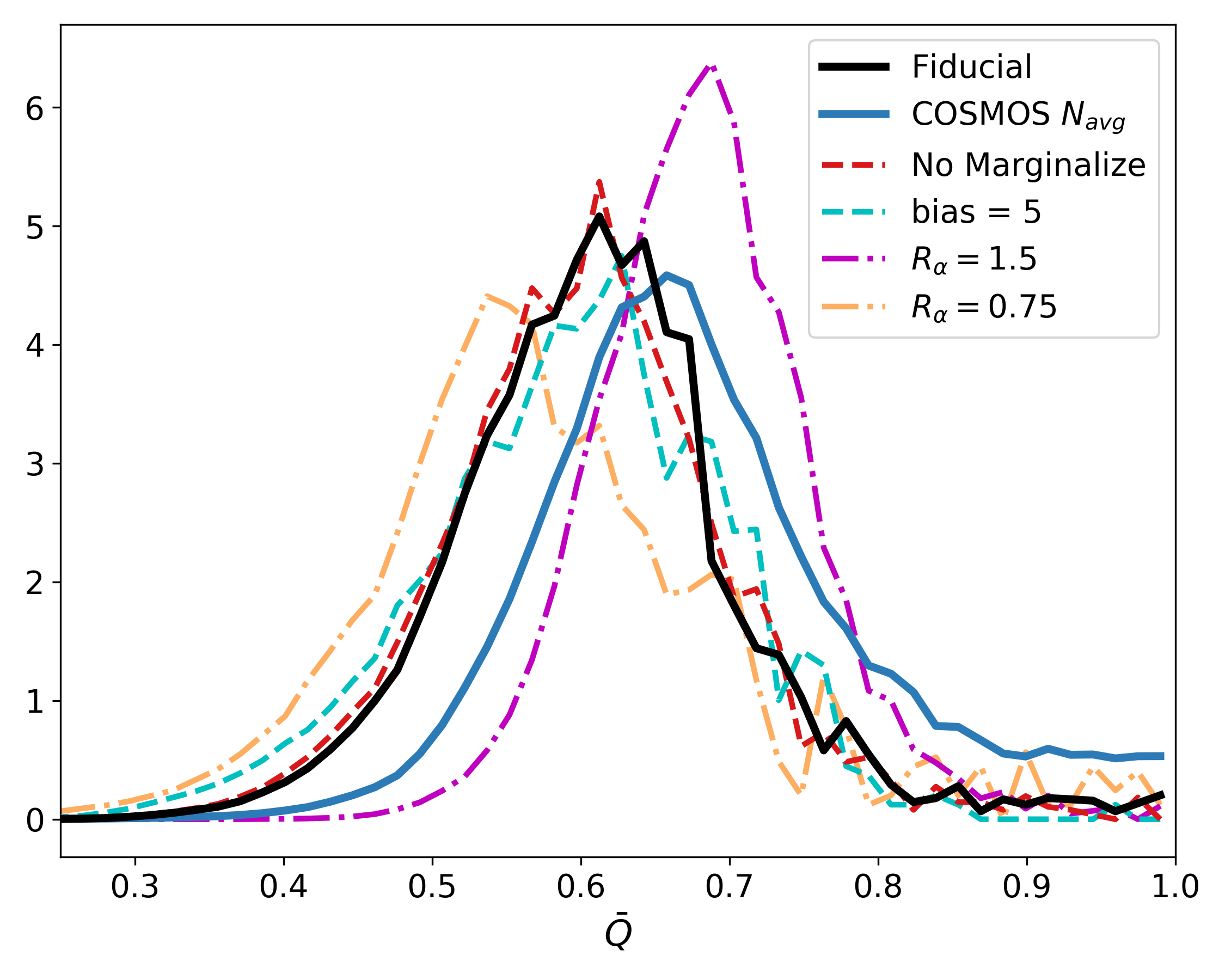}
    \caption{
    Exploring the sensitivity of our results to model choices.
    The bias value $b$ does not have a strong effect on the inference of the average ionization $\bar{Q}$. The choice of $R_\alpha$ has a larger effect on the results, and the observational parameter $N_\textrm{avg}$ has a similar impact.}
    \label{lae_fig:validation}
\end{figure}

In Figure~\ref{lae_fig:validation}, we explore the sensitivity of our results to model choices by re-running our inference in the following cases:
\begin{enumerate}
    \item COSMOS $N_\textrm{avg}$: Using the COSMOS field alone to calculate the expected number of sources in the \citet{Hu2021} volume, $N_\textrm{avg} = 2.2 \pm  0.3$. The posterior on the ionization moves to $\bar{Q} = 0.66^{+0.12}_{-0.09}$. The density posterior becomes $\delta = 0.15 \pm 0.05$
    \item No Marginalize: when re-doing our inference without marginalizing over the uncertainty in $\Nvis$ nor $b$, the posterior on the ionization stays the same, $\bar{Q} = 0.60^{+0.08}_{-0.09}$. The density posterior becomes $\delta = 0.19 \pm 0.05$.
    \item bias $= 5$: when changing the bias values from $b =$ 7.31 to 5 (with same relative uncertainty), the posterior on the ionization moves to $\bar{Q} = 0.59^{+0.09}_{-0.10}$. The density posterior stays the same at $\delta = 0.18 \pm 0.05$.
    \item $R_\alpha =$ 1.5~Mpc: when increasing the parameter $R_\alpha$, the posterior on the ionization moves to $\bar{Q} = 0.67^{+0.07}_{-0.07}$. The density posterior becomes $\delta = 0.17 \pm 0.05$.
    \item $R_\alpha =$ 0.75~Mpc: when reducing the parameter $R_\alpha$, the posterior on the ionization moves to $\bar{Q} = 0.56^{+0.12}_{-0.09}$. The density posterior stays the same at $\delta = 0.18 \pm 0.05$.
\end{enumerate}
The bias value $b$ does not appear to have a strong effect on the inference of the \citet{Hu2021} region's density nor the average ionization $\bar{Q}$. The choice of $R_\alpha$ has a larger effect on the results, and the observational parameter $N_\textrm{avg}$ has a similarly-sized impact.
\iffalse
We explore the sensitivity of our results to model choices by re-running our inference in the following cases:
\begin{enumerate}
    \item No marginalization: when re-doing our inference without marginalizing over the uncertainty in $\Nvis$ nor $b$, the posterior on the ionization moves to $\bar{Q} = 0.69^{+0.13}_{-0.09}$. The density posterior becomes $\delta = 0.14 \pm 0.05$.
    \item Reduced bias: when changing the bias values from $b =$ 7.31 to 5 (with same relative uncertainty), the posterior on the ionization moves to $\bar{Q} = 0.70^{+0.12}_{-0.09}$. The density posterior becomes $\delta = 0.14 \pm 0.05$.
    \item $R_\alpha =$ 0.75~Mpc: when reducing the parameter $R_\alpha$, the posterior on the ionization moves to $\bar{Q} = 0.67^{+0.17}_{-0.11}$. The density posterior does not change at all, staying at $\delta = 0.14 \pm 0.05$.
    \item $R_\alpha =$ 1.5~Mpc: when increasing the parameter $R_\alpha$, the posterior on the ionization moves to $\bar{Q} = 0.76^{+0.10}_{-0.07}$. The density posterior does not change at all, staying at $\delta = 0.14 \pm 0.05$.
\end{enumerate}
$\Nvis$ and the bias value $b$ do not appear to have a strong effect on the inference of the \citet{Hu2021} region's density nor the average ionization $\bar{Q}$. The choice of $R_\alpha$ has a larger effect on the results, though not dramatically.
\fi

In equation~(\ref{lae_eq:Qzetafcoll}), we assume all galaxies contribute equally to reionization. A more realistic model of reionization would change the mapping between $\bar{Q}$ and $\Fsurvavg$ (see Fig.~\ref{lae_fig:Fsurvive}). However, this change might only have a small effect on our results, given that they do not vary strongly with the choice of bias, which modifies the mapping between $\bar{Q}$ and $\Fsurvavg$ significantly. A more realistic model of reionization would likely have more of an effect on our results through $\Fsurv$, the distribution of which is important in our model and is evaluated on a small scale that is thus more subject to potentially complex behaviors of reionization. For example, analytic models of reionization tend to underestimate the bubble sizes in three-dimensional simulations \citep{Lin2016}, which we expect will allow LAEs to remain visible at earlier times. However, a comparison with the results of \citet{Mesinger2008} suggests this is not a dramatic effect. 

In our MC simulation, we included a treatment of the ``look-elsewhere'' effect. That is, the \citet{Hu2021} volume was chosen out of a much larger volume because it had the most sources. This could potentially bias the inference of the region's density, ionization, etc.
(especially because this volume was selected ``by eye" rather than in a blind tiling of the survey volume). If we remove the treatment by choosing $N_{\textrm{vols}} = 1$ in \textit{(ii)} of section~\ref{lae_sec:MC} instead of $N_{\textrm{vols}} = 118$, the peak in the likelihood for $\bar{Q}$ is shifted significantly from $\sim$0.4 to less than 0.25. However, because our prior rules out those values, the resulting effect on the posterior is very small, shifting it to $\bar{Q} = 0.59^{+0.09}_{-0.10}$.
Thus the most crucial aspect of our model is likely the prior on the underlying galaxy density, which is essential to interpreting the $\Fsurv$ values.

%%%%%%%%%%%%%%%%%%%%%%%%%%%%%%%%%%%%%
\subsection{How important are extreme regions?}

One key question with our method is how strongly it leverages extreme LAE associations in order to constrain the underlying parameters. To address this, we perform two simple tests. 

\begin{figure}
    \centering
    \includegraphics[width=0.485\textwidth]{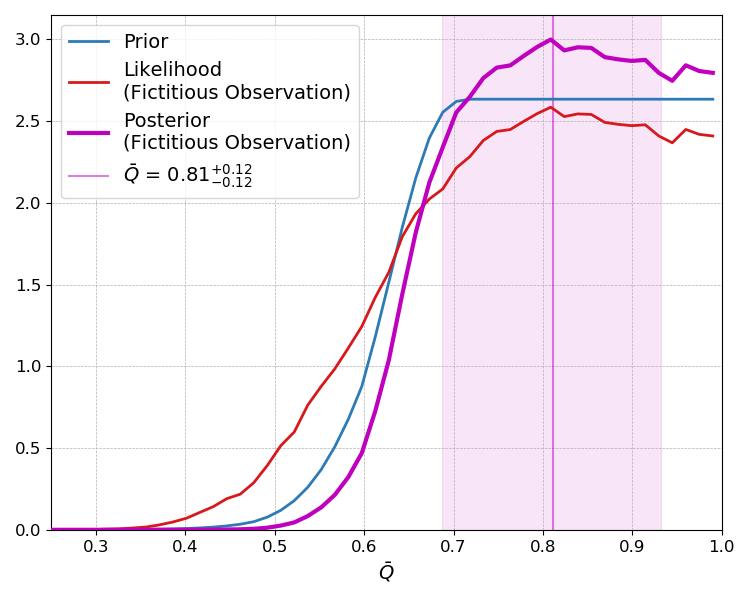}
    \caption{Constraints on the global ionized fraction $\bar{Q}$ from the \citet{Hu2021} observation, for an imaginary observation in which 7 sources were found instead of 14. The \textit{magenta} curve shows the posterior on the ionization of the Universe and 68.27\% credible interval $\bar{Q} = 0.81^{+0.12}_{-0.12}$. 
    The \textit{red} curve shows the likelihood of finding 7 sources given the average ionization of the Universe $\bar{Q}$ at $z =$ 6.93 and other model parameters, while the 
    \textit{blue} curve shows the prior on the ionized fraction of the Universe, assuming the intrinsic number of LAEs increased from $z=$ 6.93 to 5.7.
    Little information is gained in this imaginary example, as the posterior looks similar to the prior. Thus, it is only in the most extreme environments where we can learn about the average ionization of the Universe using this method.
    }
    \label{lae_fig:QbarPost_N7}
\end{figure}

First, we re-do our analysis of the \citet{Hu2021} volume, imagining that only 7 LAEs were found instead of 14. This still represents a large $\sim 4 \times$ over-density, but the results change significantly. Figure~\ref{lae_fig:QbarPost_N7} shows the posterior of the globally-averaged ionized fraction shifts from $\bar{Q} = 0.60^{+0.08}_{-0.09} \rightarrow 0.81^{+0.12}_{-0.12}$, and becomes more like a lower-limit, barely adding information to the prior.
This counts-in-cells framework therefore gains the most information from the most extreme environments. A modest over-density only provide modest limits, although we have not tested the extent to which the full distribution of counts-in-cells can add information.

%%%%%%%%%%%%%%%%%%%%%%%%%%%%%%%%%%%%%
%\subsection{Results for a different over-density}

\begin{figure}
    \centering
    \includegraphics[width=0.485\textwidth]{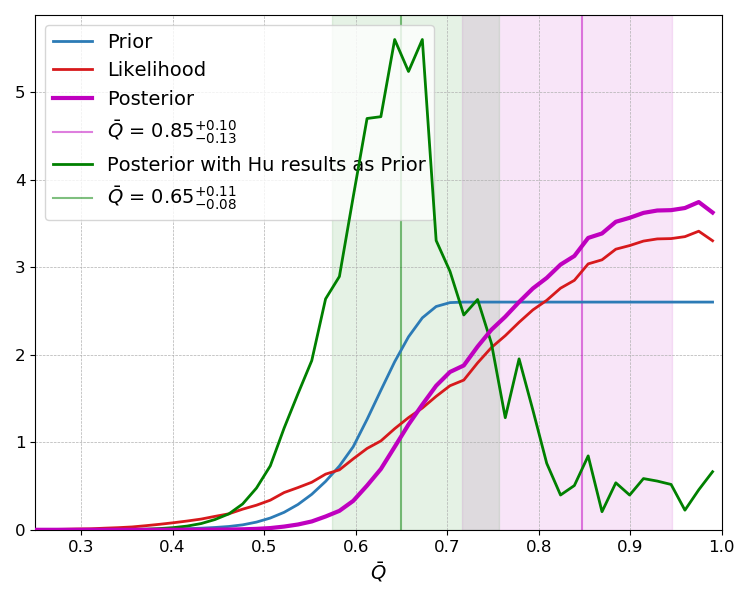}
    \caption{Constraints on the  $z =$ 6.8 global ionized fraction $\bar{Q}$ from the \citet{Endsley2022} volume, where 6 sources were found expecting only 2. The \textit{magenta} curve shows the posterior on the ionization of the Universe and 68.27\% credible interval $\bar{Q} = 0.85^{+0.10}_{-0.13}$. 
    The \textit{red} curve shows the likelihood of finding 6 sources given the average ionization of the Universe $\bar{Q}$ at $z =$ 6.8 and other model parameters, while the 
    \textit{blue} curve shows the prior on the ionized fraction of the Universe, assuming the intrinsic number of LAEs increased from $z=$ 6.93 to 5.7.
    The \textit{green} curve combines the likelihood with the likelihood of the \citet{Hu2021} volume (see Fig.~\ref{lae_fig:QbarPost}) and the prior, shifting the posterior to $\bar{Q} = 0.65^{+0.11}_{-0.08}$. %These measurements are $\Delta z \simeq 0.1$ apart from one another, over which time the Universal ionization state changed much less than the uncertainty in our measurements. 
    }
    \label{lae_fig:QbarPost_Endsley}
\end{figure}

As a second test, we apply our method to the \citet{Endsley2022} volume, where 6 sources were found expecting only 2 (see section~\ref{lae_sec:Endsley}). Figure~\ref{lae_fig:QbarPost_Endsley} shows the results. 
These data can constrain the $z =$ 6.8 global ionized fraction $\bar{Q}$ to 68.27\% credible interval $\bar{Q} = 0.85^{+0.10}_{-0.13}$. The \textit{red} curve shows the likelihood of finding 6 sources given the average ionization of the Universe $\bar{Q}$ at $z =$ 6.8 and other model parameters, while the 
\textit{blue} curve shows the prior on the ionized fraction of the Universe, assuming the intrinsic number of LAEs increased from $z=$ 6.93 to 5.7.
The \textit{green} curve combines the likelihood with the likelihood of the \citet{Hu2021} volume (see Fig.~\ref{lae_fig:QbarPost}) and the prior, shifting the posterior to $\bar{Q} = 0.65^{+0.11}_{-0.08}$. These measurements are $\Delta z \simeq 0.1$ apart from one another, over which time the universal ionization state changed much less than the uncertainty in our measurements.

%%%%%%%%%%%%%%%%%%%%%%%%%%%%%%%%%%%%%%%%%%%%%%%%%%%%%%%%%%%%%%%%
%%%%%%%%%%%%%%%%%%%%%%%%%%%%%%%%%%%%%%%%%%%%%%%%%%%%%%%%%%%%%%%%

%%%%%%%%%%%%%%%%%%%%%%%%%%%%%%%%%%%%%%%%%%%%%%%%%%%%%%%%%%%%%%%%
%%%%%%%%%%%%%%%%%%%%%%%%%%%%%%%%%%%%%%%%%%%%%%%%%%%%%%%%%%%%%%%%
\section{Conclusions}\label{lae_sec:discussion}

%Summary of problem(s): density estimates, ionization field from galaxy/LAE associations BUT only seeing the tip of the iceberg
There is a long history of close study of ``extreme" objects in astrophysics, as they can offer sharp tests of our physical paradigms. At high redshifts, associations of bright galaxies are often labeled as ``protoclusters,'' although the mapping to present-day clusters has only been qualitative. Because the neutral IGM modulates Lyman-$\alpha$ absorption, associations of LAEs have similarly been suspected of identifying large ionized regions. 
In this paper, we have introduced a framework to make these identifications in a statistically rigorous manner. We first construct an analytic form of the posterior for the underlying density of a region given a finite number of observed galaxies, ignoring any modulation from reionization. This framework combines the effects of cosmic variance and Poisson noise, making use of a strong prior on the density of a region. 

Inferences about the ionization field require a model of the reionization process. Here we have used a simple model motivated by the recent measurement of a surprisingly short mean free path for ionizing photons during this era \citep{Becker2021}. Using this model in a Monte Carlo simulation, we found that sufficiently large LAE associations can not only identify ionized volumes with high reliability but also constrain reionization on a global scale. 

Assuming a fully-ionized Universe at $z =$ 6.93, we calculate the linear dark-matter density of the \citet{Hu2021} volume to be $\delta = 0.19 \pm 0.06$. When considering a partially-ionized Universe via a simple model of re-ionization, we calculate the density to be slightly lower, $\delta = 0.18 \pm 0.05$. These densities are $\sim2\sigma$ below the required linear density of $\delta_{\textrm{pc}} = 0.27$ for this region to collapse into a single virialized object by $z = $ 0.
The \citet{Hu2021} volume gives a constraint on the ionized fraction of the Universe at $z =$ 6.93, $\bar{Q} = 0.60^{+0.08}_{-0.09}$. This result is strongly prior-dominated at low values of $\bar{Q}$, and should only be interpreted in the context of our very simple model of reionization; the quoted errors do not include systematic uncertainties in the reionization model.
We constrain the ionized fraction of the \citet{Hu2021} volume itself to be $Q > 0.74$ at 95.45\% credibility.

Our inferences about reionization are subject to systematic uncertainties about the underlying reionization model, but we already find that even a \emph{single} well-defined LAE association offers competitive constraints on the global ionized fraction at $z \sim 7$ \citep{Greig2017,Inoue2018,Mason2018,Davies2018}; our results are consistent with other methods and have similar uncertainties. 
%Moreover, it has different systematic uncertainties than other approaches, even those that rely on evolution in the number density or large-scale clustering of LAEs. 
In contrast to most inferences from LAEs, our approach is more similar to a ``counts-in-cells" method that leverages the non-gaussianity that reionization induces in the LAE distribution. 

%Emphasize provides framework to infer local properties (density, ionization), improvement over qualitative efforts that have so far been made. Can then use these locally-understood regions as targets for studies of effects of reionization on faint galaxies etc.
An advantage of our framework is that it identifies ionized regions in well-specified locations on the sky -- providing targets for detailed studies of the effect of these ionized regions on the galaxy populations.

The constraints from the simple exercise in this paper suggest that the counts-in-cells approach may be very powerful. We have focused on a single association using a simple model. Future improvements to this framework could include: \textit{(i)} a more complete reionization/Lyman-$\alpha$ absorption model, \textit{(ii)} incorporating the distribution of source luminosities with a LAE luminosity function, \textit{(iii)} considering all observed regions simultaneously rather than just a single association (or in other words implementing a full counts-in-cells framework), \textit{(iv)} simultaneously incorporating information from photometric galaxy selection (e.g., \citealt{Yoshioka2022}) and LAE surveys, and \textit{(v)} considering the expansion/contraction of a region depending on its density, which increases the relative odds of finding an under-dense region \citep{Munoz2010,Trapp2020}.

Future observations with JWST and other telescopes will discover many LAEs at even higher redshifts. This hugely increased sample -- combined with a more accurate model of reionization -- is a promising avenue for calculating the ionized fraction of the Universe throughout reionization.

%%%%%%%%%%%%%%%%%%%%%%%%%%%%%%%%%%%%%%%%%%%%%%%%%%%%%%%%%%%%%%%%
%%%%%%%%%%%%%%%%%%%%%%%%%%%%%%%%%%%%%%%%%%%%%%%%%%%%%%%%%%%%%%%%

%%%%%%%%%%%%%%%%%%%%%%%%%%%%%%%%%%%%%%%%%%%%%%%%%%%%%%%%%%%%%%%%
%%%%%%%%%%%%%%%%%%%%%%%%%%%%%%%%%%%%%%%%%%%%%%%%%%%%%%%%%%%%%%%%
\section*{Acknowledgements}

We thank S.~Naoz, A.~Shapley, and T.~Treu for helpful discussions. This work was supported by the National Science Foundation through award AST-1812458. In addition, this work was directly supported by the NASA Solar System Exploration Research Virtual Institute cooperative agreement number 80ARC017M0006. We also acknowledge a NASA contract supporting the ``WFIRST Extragalactic Potential Observations (EXPO) Science Investigation Team'' (15-WFIRST15-0004), administered by GSFC. 

%%%%%%%%%%%%%%%%%%%%%%%%%%%%%%%%%%%%%%%%%%%%%%%%%%%%%%%%%%%%%%%%
%%%%%%%%%%%%%%%%%%%%%%%%%%%%%%%%%%%%%%%%%%%%%%%%%%%%%%%%%%%%%%%%

%%%%%%%%%%%%%%%%%%%%%%%%%%%%%%%%%%%%%%%%%%%%%%%%%%%%%%%%%%%%%%%%
%%%%%%%%%%%%%%%%%%%%%%%%%%%%%%%%%%%%%%%%%%%%%%%%%%%%%%%%%%%%%%%%
\section*{Data Availability}

There are no novel data produced in this work.

%%%%%%%%%%%%%%%%%%%%%%%%%%%%%%%%%%%%%%%%%%%%%%%%%%%%%%%%%%%%%%%%
%%%%%%%%%%%%%%%%%%%%%%%%%%%%%%%%%%%%%%%%%%%%%%%%%%%%%%%%%%%%%%%%



\bibliographystyle{mnras}
\bibliography{me} % if your bibtex file is called example.bib

%\begin{thebibliography}{99}
%\end{thebibliography}

%%%%%%%%%%%%%%%%%%%%%%%%%%%%%%%%%%%%%%%%%%%%%%%%%%

%%%%%%%%%%%%%%%%% APPENDICES %%%%%%%%%%%%%%%%%%%%%

%\appendix
%%%%%%%%%%%%%%%%%%%%%%%%%%%%%%%%%%%%%%%%%%%%%%%%%%

% Don't change these lines
\bsp	% typesetting comment
\label{lastpage}
\end{document}